\documentclass{article}

\usepackage{amsfonts, amsmath, amssymb}
\usepackage{setspace}
\usepackage{graphicx}
\usepackage{psfrag}
\usepackage{graphics}
\usepackage{subfigure}

\usepackage{soul,color}
\usepackage{marginnote}

\usepackage{datetime}

\graphicspath{{Figures/}}

\newcommand{\W}{\mathcal{W}}
\newcommand{\C}{\mathcal{C}}

\newcommand{\di}{{\rm d}}

\begin{document}

\title{Earth--Mars Transfers with Ballistic Capture}

\author{\scshape Edward Belbruno${}^1$ \hspace{.5cm} Francesco Topputo${}^2$}

\date{}

\maketitle

\vspace{-5mm}

{
\footnotesize
\centerline{${}^1$  Princeton University, Princeton, New Jersey 08544, USA}
\centerline{${}^2$  Politecnico di Milano, Milan 20156, Italy}
}

\maketitle

\vspace{5mm}

\begin{abstract}

\noindent We construct a new type of transfer from the Earth to Mars, which ends in ballistic capture. This results in a substantial savings in capture $\Delta v$ from that of a classical Hohmann transfer under certain conditions. This is accomplished by first becoming captured at Mars, very distant from the planet, and then from there, following a ballistic capture transfer to a desired altitude within a ballistic  capture set. This is achieved by manipulating the stable sets, or sets of initial conditions whose orbits satisfy a simple definition of stability. This transfer type may be of interest for Mars missions because of lower capture $\Delta v$, moderate flight time, and flexibility of launch period from the Earth.

\end{abstract}

\section{Introduction} \label{sec:intro}

In 1991 the Hiten spacecraft of Japan used a new type of transfer to the Moon, using ballistic capture \cite{belbruno1993}.  This is a capture where the Kepler energy of the spacecraft with respect to the Moon becomes negative from initially positive values, by only using the natural gravitational forces of the Earth, Moon and Sun. It is generally temporary. This capture uses substantially less $\Delta v$ than a Hohmann transfer which has a positive $v_{\infty}$ at lunar approach, making it an attractive alternative for lunar missions. This same type of transfer was, in fact, used by NASA's GRAIL mission in 2011 \cite{Chung2010}. Another type of ballistic capture transfer first found in 1986, was used in 2004 by ESA's SMART-1 mission \cite{belbruno2007,belbruno2004}.

Since ballistic capture occurs about the Moon in a region called a weak stability boundary, these transfers are called weak stability boundary transfers or ballistic capture transfers. The types that were used for Hiten and GRAIL are called {\em exterior} transfers since they first go beyond the orbit of the Moon. The types used for SMART-1 are called {\em interior} transfers since they remain within the Earth--Moon distance \cite{belbruno2004}.  They are also referred to as low energy transfers, since they use less $\Delta v$ for capture. The weak stability boundary, in general,  has recently been shown to be a complex fractal region consisting of a network of invariant manifolds, associated to the collinear Lagrange points, $L_1, L_2$  \cite{belbruno2004,belbruno2013,belbruno2010}. The dynamics of motion in this region is chaotic and unstable, thus explaining why the capture is temporary. 

Ever since these ballistic capture transfers to the Moon were discovered, it was natural to ask if there were transfers from the Earth that led to ballistic capture at Mars. It was generally felt that Hiten-like transfers did not exist after a number of efforts \cite{lo1998,castillo2003,topputo2005b,mingotti2011c}. The reason for this is that the orbital velocity of Mars is much higher than the approach $v_{\infty}$ of a Hohmann transfer from the Earth, whereas the $v_{\infty}$ of a Hohmann transfer to the Moon is close to the Moon's orbital velocity.

The purpose of this paper is to show that ballistic capture transfers to Mars, from the Earth, do exist. We will show how to construct them. The key idea is not to try to find transfers from the Earth that go directly to ballistic capture near to Mars. But rather, to first transfer to ballistic capture far from Mars, many millions of kilometers away from Mars, yet close to its orbit about the Sun. At first it would seem counter intuitive to first transfer so far from Mars. At this distant location, ballistic capture transfers can be found that go close to Mars after several months travel time,  in the examples given, and then into ballistic capture. This results in elliptic-type orbits about Mars. We show that for periapsis altitudes higher than 22,000 km, these transfers from the Earth use considerably less $\Delta v$ than a Hohmann transfer.   At altitudes less than this, say 100 km, it is found that the Hohmann transfer uses only slightly less capture $\Delta v$ which may make the ballistic capture alternative presented here more desirable. This is because by transferring from the Earth to points far from Mars near Mar\rq{}s orbit, it is not necessary to adhere to a  2 year launch period from the Earth. The times of launch from the Earth can be much more flexible. 

The use of this new transfer may have a number of advantages for Mars missions.  This includes substantially lower capture $\Delta v$ at higher altitudes, flexibility of launch period from the Earth, gentler capture process, first transferring to locations far from Mars offering interesting new approaches to Mars itself, being ballistically captured into capture ellipses for a predetermined number of cycles about Mars, and the ability to transfer to lower altitudes with relatively little penalty.  The initial capture locations along Mars orbit may be of interest for operational purposes. 

The structure of this paper is as follows: In Section \ref{sec:1}, we describe the methodology and steps that we will use to find these new transfers.   In the remaining sections, these steps are elaborated upon. In Section \ref{sec:2}, we describe the basic model used to compute the trajectories, planar elliptic restricted three-body problem. In Section \ref{sec:3}, the stable sets at Mars are described, whose manipulation allows us to achieve the capture sets. In Section \ref{sec:4} we describe interplanetary transfers from Earth to locations far from Mars that are at the beginning of ballistic capture transfers to Mars. In Section \ref{sec:5} comparisons to Hohmann transfers are made. In Section \ref{sec:6} applications are discussed and future work. Two Appendixes are reported where complementary material is presented.

\section{Methodology and Steps} \label{sec:1}

The new class of ballistic capture transfers from Earth to Mars are constructed in a number of steps. These steps are as follows:
\medskip

\noindent
{\em Step 1} --- Compute a ballistic capture trajectory (transfer) to Mars to a given periapsis distance, $r_p$, that starts far from Mars at a point, $\bf x_c$ near Mars orbit.  In this paper $\bf x_c$ is arbitrarily chosen several million kilometers from Mars. $\bf x_c$ corresponds to the start of a trajectory that goes to ballistic capture near Mars, after a maneuver,  $\Delta V_c$, is applied (defined in the next step). Although this location is far from Mars, we refer to it as a capture maneuver, since the trajectory eventually leads to ballistic capture. It takes, in general, several months to travel from $\bf x_c$ to ballistic capture near Mars at a periapsis distance, $r_p$. When it arrives at the distance $r_p$, its osculating eccentricity, $e$, with respect to Mars is less then $1$.  Once the trajectory moves beyond the capture at the distance $r_p$, it is in a special capture set where it will perform a given number of orbits about Mars.  The simulations in this step use the planar elliptic restricted three-body problem.
\medskip

\noindent
{\em Step 2} --- An interplanetary transfer trajectory for the spacecraft, $P$, starts at the SOI of the Earth. A maneuver, $\Delta V_1$,  is applied to transfer to the point $\bf x_c$ near Mars orbit, where a maneuver, $\Delta V_c$, is used to match the velocity of the ballistic capture transfer to Mars. This transfer is in heliocentric space and is viewed as a two-body  problem between $P$ and the Sun.  $\Delta V_1$,  $\Delta V_c$ are minimized. 
\medskip

\noindent
{\em Step 3} --- The trajectory consisting of the interplanetary transfer to $\bf x_c$ together with the ballistic capture transfer from $\bf x_c$ to the distance $r_p$ from Mars (with osculating eccentricity $e < 1$) is the resulting ballistic capture transfer from the Earth. This is compared to a standard Hohmann transfer leaving the Earth from the same distance, in the SOI, and going directly to the distance $r_p$ from Mars with the same eccentricity $e$, where a $\Delta V_2$ is applied at the distance $r_p$ to achieve this eccentricity. $\Delta V_2$  is compared to $\Delta V_c$. It is found in the cases studied, that for $r_p > 22,000$km,  we can achieve $\Delta V_c < \Delta V_2$ .  It is found that $\Delta V_c$ can be on the order of $25\%$  less then $\Delta V_2$ if the value of $r_p$ is approximately $200,000$km. It is shown that by transferring to much lower altitudes from these $r_p$ values yields only a relatively small increase from the capture $\Delta v$ required for a Hohmann transfer.  As is explained in latter sections, this may make the ballistic capture transfer more desirable in certain situations.

\begin{figure}
\centering
	\includegraphics[width=0.9\textwidth, clip, keepaspectratio]{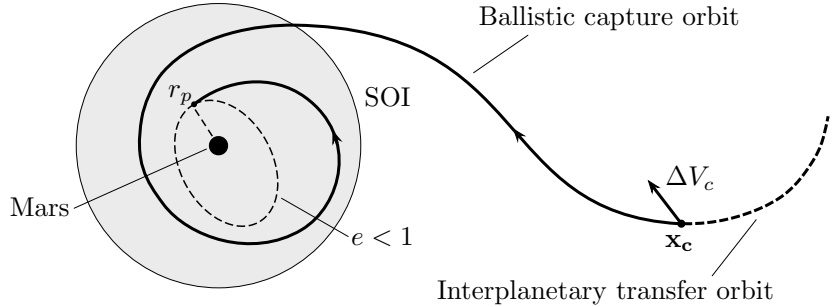}
	    \caption{Structure of the ballistic capture transfers to Mars.}
    \label{fig:TransferStructure}
\end{figure}

The main reasons $\bf x_c$ is chosen far from Mars is three-fold. First, if $\bf x_c$ is sufficiently far from the Mars SOI, there is negligible gravitational attraction of Mars on $P$. This yields a more constant arrival velocity from the Earth in general. Second, since the points, $\bf x_c$, lie near to Mar\rq{}s orbit, there are infinitely many of them which offer many locations to start a ballistic capture transfer. This variability of locations gives flexibility of the launch period from the Earth. Third, since $\bf x_c$ is outside the SOI of Mars, the application of $\Delta V_c$ can be done in a gradual manner, and from that point on, no more maneuvers are required, where $P$ arrives at the periapsis distance $r_p$ in a natural capture state.  This process is much more benign that the high velocity capture maneuver at $r_p$ that must be done by a Hohmann transfer. From an operational point of view, this is advantageous.

\smallskip

We now describe these steps in detail in the following sections.

\section{Model} \label{sec:2}

When our spacecraft, $P$,  is in motion about Mars, from arrival at $\bf x_c$ to Mars ballistic capture at $r_p$, we model the motion of $P$ by the planar elliptic restricted three-body problem, which takes into account Mars eccentricity $e_p = 0.093419$. We view the mass of $P$ to be zero. 

The planar elliptic restricted three-body problem studies the motion of a massless particle, $P$, under the gravitational field generated by the mutual elliptic motion of two primaries, $P_1$, $P_2$, of masses $m_1$, $m_2$, respectively. In this paper, $P_1$ is the Sun, and $P_2$ is Mars. The equations for the motion of $P$ are
\begin{equation} \label{eq:BasicDEs}
	x'' - 2 y' = \omega_{x}, \qquad y'' + 2 x' = \omega_{y}.
\end{equation}
The subscripts in Eq.\ \eqref{eq:BasicDEs} are the partial derivatives of
\begin{equation} \label{eq:omega}
	\omega(x,y,f) = \frac{\Omega(x,y)}{1 + e_p \cos f},
\end{equation}
where the potential function is
\begin{equation} \label{eq:Omega}
	\Omega(x,y) = \frac{1}{2}(x^2 + y^2) + \frac{1 - \mu}{r_1} + \frac{\mu}{r_2} + \frac{1}{2}\mu(1 - \mu),
\end{equation}
and $r_1 = \left[(x + \mu)^2 + y^2\right]^{1/2}$, $r_2 = \left[(x + \mu - 1)^2 + y^2\right]^{1/2}$.

\smallskip

Equations \eqref{eq:BasicDEs} are written in a nonuniformly rotating, barycentric, adimensional coordinate frame where $P_1$ and $P_2$ have fixed positions $(-\mu, 0)$ and $(1-\mu, 0)$, respectively, and $\mu = m_2/(m_1+m_2)$ is the mass parameter of the system, $\mu = 3.2262081094\times10^{-7}$.  This coordinate frame isotropically pulsates as the $P_1$--$P_2$ distance, assumed to be the unit distance, varies according to the mutual position of the two primaries on their orbits (see \cite{szebehely1967} for the derivation of Eqs.\ \eqref{eq:BasicDEs}). The primes in Eq.\ \eqref{eq:BasicDEs} represent differentiation with respect to $f$, the true anomaly of the system. This is the independent variable, and plays the role of the time: $f$ is assumed to be zero when $P_1$, $P_2$ are at their periapse, as both primaries orbit the center of mass in similarly oriented ellipses having common eccentricity $e_p$. Normalizing the period of $P_1$, $P_2$ to $2\pi$, the dependence of true anomaly on time, $t$, 
\begin{equation} \label{eq:trueanomaly}
	f(t) = f_0 + \int_{t_0}^t \frac{(1 + e_p \cos f(\tau))^2}{(1 - e_p^2)^{3/2}}\ \di\tau,
\end{equation}
where $f_0$ and $t_0$ are the initial true anomaly and time, respectively.

The elliptic problem possesses five equilibrium points, $L_k$, $k = 1,\dots,5$. Three of these, $L_1, L_2, L_3$, lie along the $x$-axis ($L_1$ lies between $P_1$ and $P_2$); the other two points, $L_4, L_5$, lie at the vertices of two equilateral triangles with common base extending from $P_1$ to $P_2$. These points have fixed location in the rotating, scaled frame. However, their real distance from $P_1$, $P_2$ varies (pulsates) according to the mutual motion of the primaries. When $e_p =0$, we obtain the planar circular restricted three-body problem.
\medskip\medskip

\section{Mars Stable Sets and Ballistic Capture Orbits} \label{sec:3}

In this section we elaborate on Step 1 in Section \ref{sec:1}. The goal is to compute special ballistic capture trajectories that start far from Mars ($P_2$) and go to ballistic capture near Mars at a specified radial distance, $r_p$. It is recalled, that a ballistic capture trajectory for $P$ with respect to $P_2$ is one where two-body (Kepler) energy of $P$ with respect to $P_2$ is initially positive and which becomes negative, where ballistic capture occurs (see \cite{belbruno2004,hyeraci2010} for more details).

Ballistic capture trajectories can be designed by making use of stable sets associated to the algorithmic definition of weak stability boundaries.

In \cite{topputo2009a}, the algorithmic definition of the WSB is given in the circular restricted three-body problem, about Jupiter, where the stable sets are computed.  These are computed by a definition of stability that can be easily extended to more complicated models. Stable sets are constructed by integrating initial conditions of the spacecraft about one primary and observing its motion as it cycles the primary, until the motion substantially deviates away from the primary. Special attention is made to those stable orbits that in backwards time, deviate before one cycle. These are good for applications for minimal energy capture. Although derived by an algorithmic definition, the dynamics  of stable sets can be related to those of the Lagrange points \cite{belbruno2010, garcia2007}, which is a deep result. 
  
More precisely, stable sets are computed by the following procedure(see \cite{topputo2009a} for more details):  A grid of initial conditions is defined around one of the two primaries in the restricted three-body problem. These correspond to periapsis points of elliptic two-body orbits with different semi-major axis and orientation. The eccentricity is held fixed in each of the stable sets. Initial conditions are integrated forward and labeled according to the stability of the orbits they generate. In particular, an orbit is deemed {\em n-stable} if it performs $n$ revolutions around the primary while having negative Kepler energy at each turn and without performing any revolution around the other primary. Otherwise, it is called {\em n-unstable}.   Backward stability is introduced by studying the behavior of the orbits integrated backward in time; this defines  $-��m$-stability. The weak stability boundary itself occurs as the boundary of the stable regions.

In the circular restricted three-body problem, the union of all $n$-stable initial conditions is indicated as $\mathcal{W}_n(e)$, where $e$ is the eccentricity used to define the initial conditions (see \cite{topputo2009a}). When computed in nonautonomous (i.e., time dependent) models, the initial conditions have to account for the initial time as well.  If the elliptic restricted three-body problem is used, the stable sets are indicated by $\mathcal{W}_n(f_0,e)$.
\medskip

The details of these definitions in the case of the elliptic restricted problem are found in \cite{hyeraci2010}. They are also summarized in the Appendix. 
\medskip

Computing stable sets involves integrating tens of thousands of orbits generated over a computational grid of points. In \cite{hyeraci2010}  polar coordinates are used, and therefore the grid is defined by radial, angular spacing of points. This shows up in the plots upon magnification. 
  
It is remarked that the set of grid points is five-dimensional. The grid is fine so not to lose relevant information about the stable sets. For this reason, the computations are time intensive. The parameters and their range and refinement are: (i.) $r$, the radial distance to Mars, spacing $\Delta r = 50$ km, for $250 \leq r \leq 30,500$ km, and $\Delta r =500$ km, for  $30,500 \leq  r \leq 250,000$ km; (ii.) $\theta$,  angular position with respect to a reference direction, $0 \leq \theta \leq 360$ deg, $\Delta \theta$ = 1 deg; (iii.) $e$, the osculating eccentricity, $0.90 \leq e \leq 0.99$, $\Delta e = 0.01$; (iv.) $f_0$, the initial true anomaly of primaries, $0 \leq f_0 \leq \pi/2$, $\Delta f_0 = \pi/4$; (v.) $n$, the stability number, $-1 \leq n \leq 6$, $\Delta n = 1$.

The spatial part of the grid, given by $\{r,\theta\}$, requires 375,394 initial conditions which need to be numerically integrated. All numerical integrations of System \ref{eq:BasicDEs} are done using a variable-order, multi-step Adams--Bashforth--Moulton scheme. Also, when $P$ comes close to Mars ($P_2$), then a Levi-Civita regularization is used to speed up the numerical integration (see \cite{topputo2009a}).

\subsection{Constructing Ballistic Capture Orbits About Mars}

In \cite{belbruno2010, hyeraci2010}, a method to construct ballistic capture orbits with prescribed stability number is given. This method is based on a manipulation of the stable sets. It is briefly recalled. First, let us consider the set $\mathcal{W}_{-1}(e, f_0)$: this set is made up of the initial conditions that generate $-1$-stable orbits; i.e., orbits that stay about the primary for at least one revolution when integrated backward. By definition, the complementary set, $\overline{\mathcal{W}}_{-1}(e, f_0)$, contains initial conditions that generate $-1$-unstable orbits. These are orbits that escape from the primary in backward times or, alternatively, they approach the primary in forward time. The ballistic capture orbits of practical interest are contained in the capture set
\begin{equation}
\mathcal{C}_{-1}^n (e, f_0) = \overline{\mathcal{W}}_{-1}(e, f_0) \cap  \mathcal{W}_n(e, f_0).
\label{eq:CapSet}
\end{equation}					
The points in $\mathcal{C}_{-1}^n$ are associated to orbits that both approach the primary and perform at least $n$ revolutions around it. This is desirable in mission analysis, as these orbits may represent good candidates to design the ballistic capture immediately upon arrival. For a proper derivation of the capture set it is important that only those sets computed with identical values of $e, f_0$ are intersected. This assures the continuity along the orbits; i.e., the endpoint of the approaching ($-1$-unstable) orbit has to correspond to the initial point of the $n$-stable orbit.

\begin{figure}[t!]
	\centering
		\subfigure[$\W_1(\pi/4,0.99)$ \label{fig:W1f025e99}]
			{\includegraphics[width=0.45\textwidth, clip, keepaspectratio]{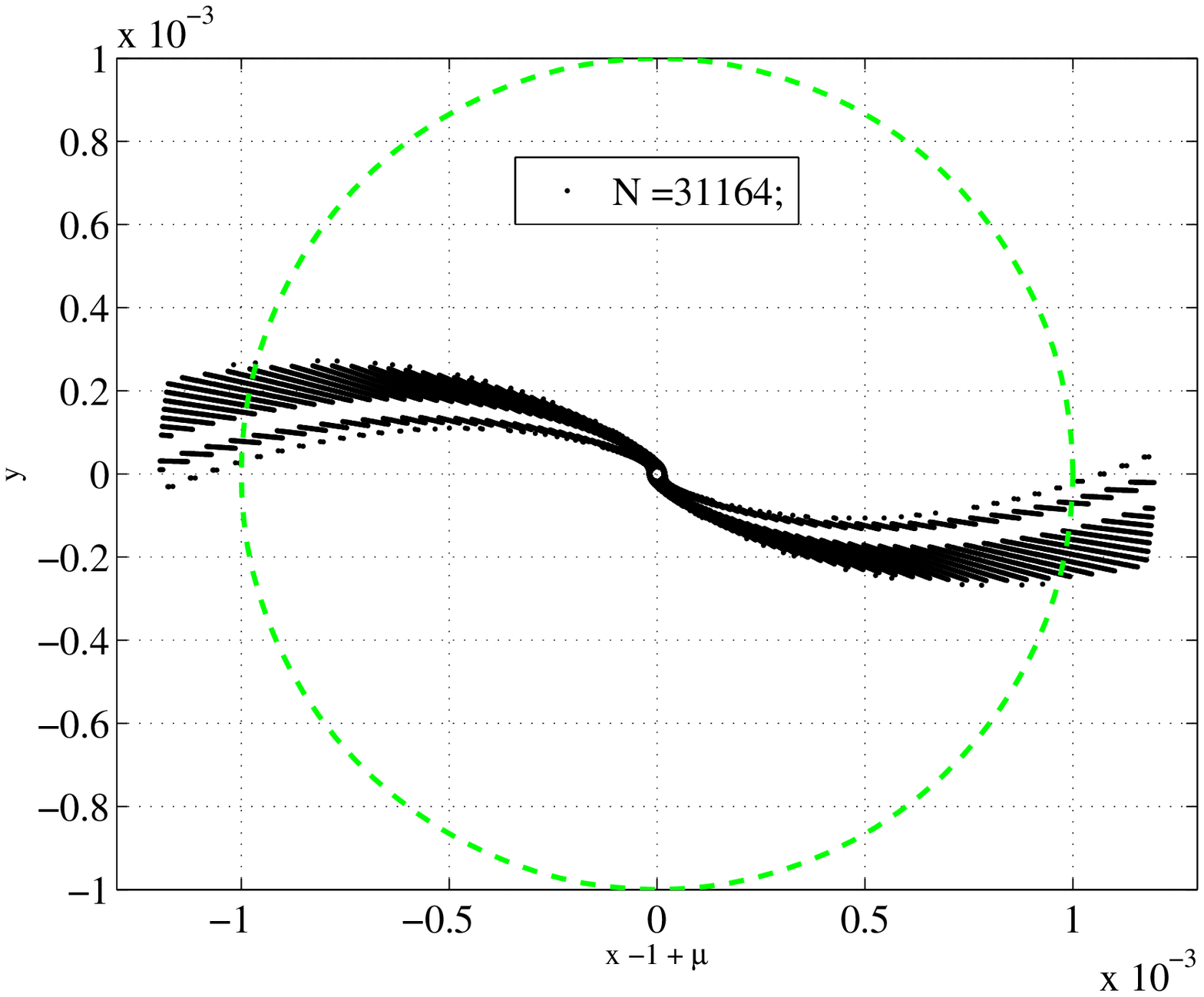}}
		\subfigure[$\W_2(0.99,\pi/4)$ \label{fig:W2f025e99}]
			{\includegraphics[width=0.45\textwidth, clip, keepaspectratio]{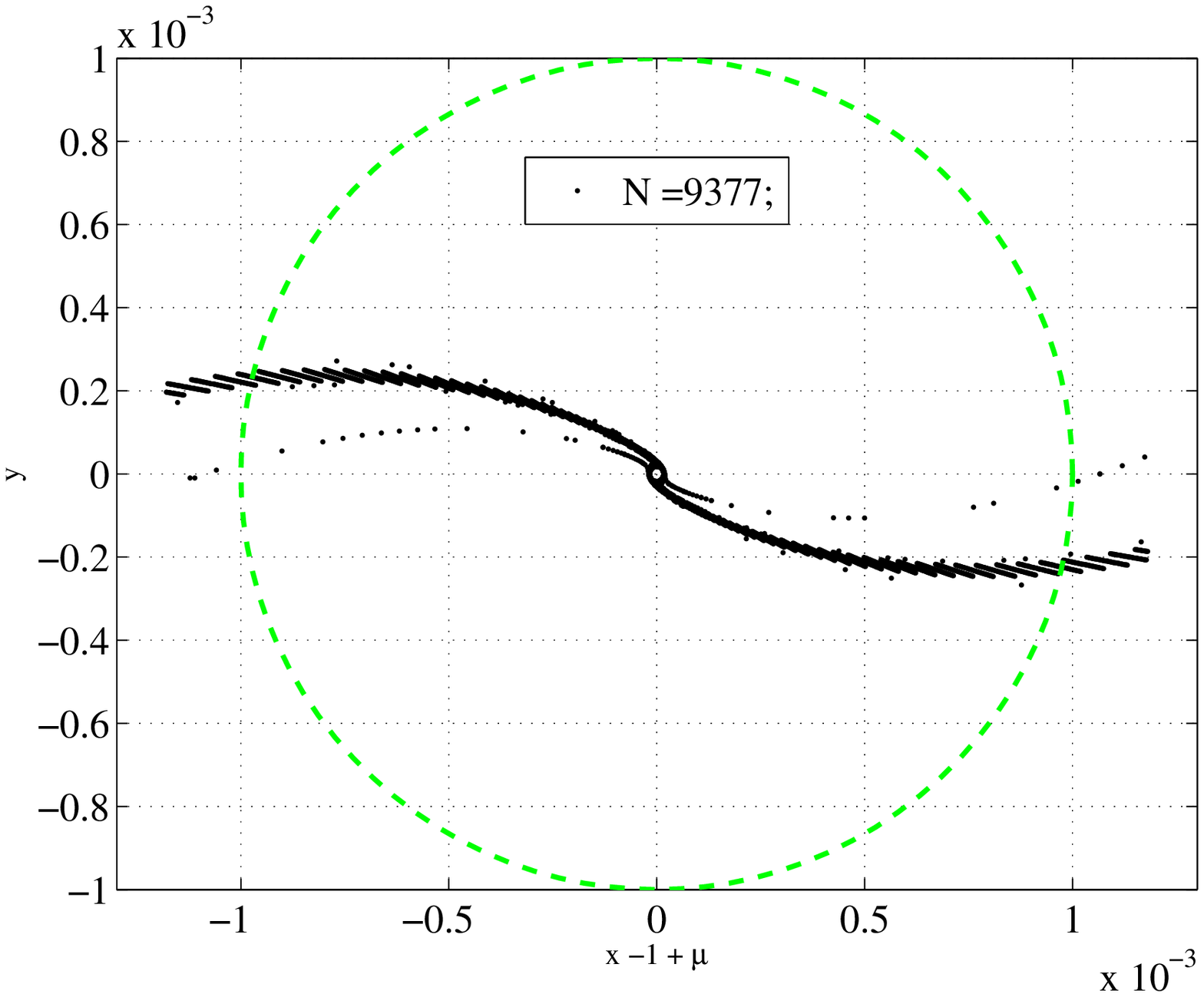}}
		\subfigure[$\W_3(0.99,\pi/4)$ \label{fig:W3f025e99}]
			{\includegraphics[width=0.45\textwidth, clip, keepaspectratio]{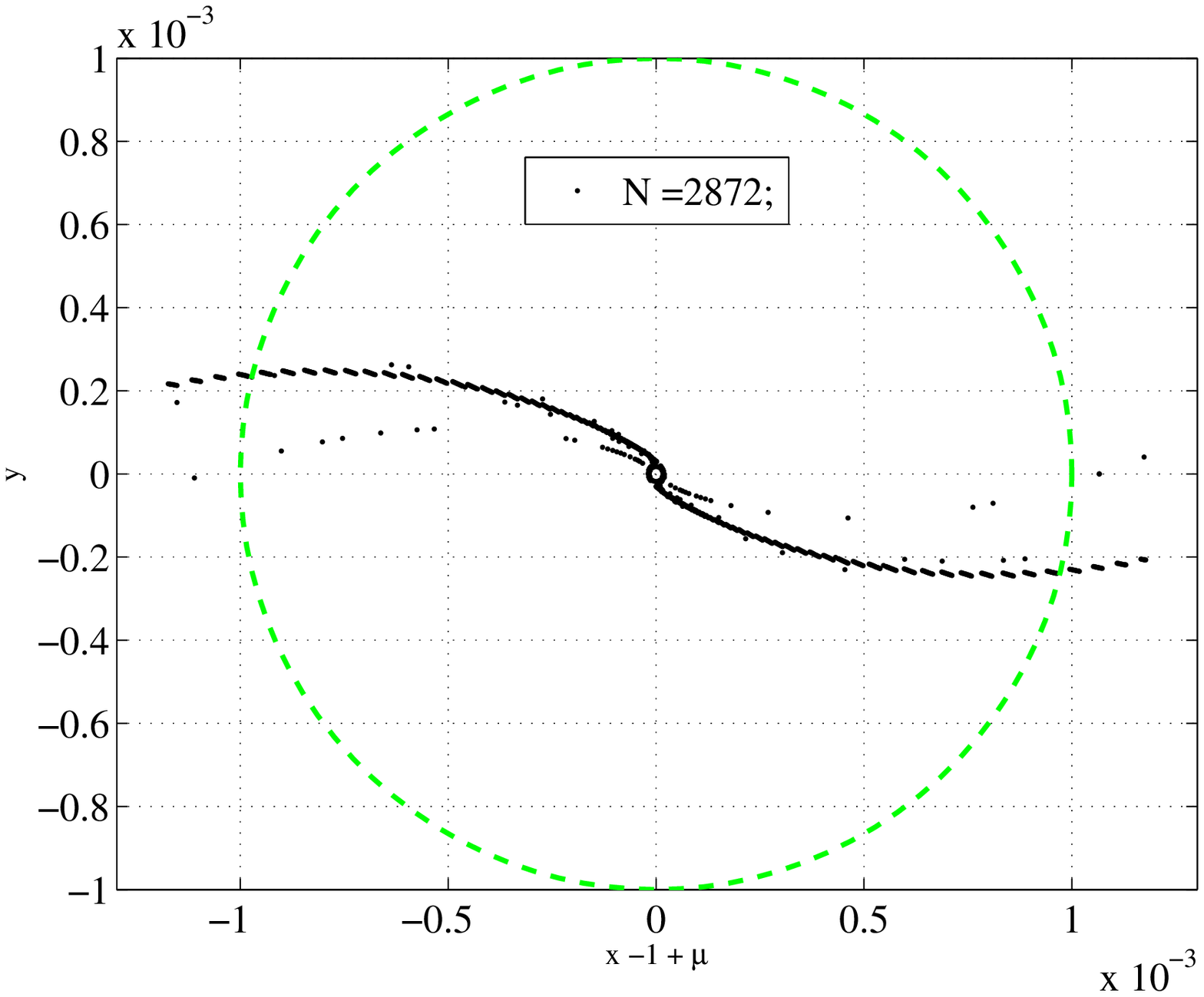}}
		\subfigure[$\W_6(0.99,\pi/4)$ \label{fig:W6f025e99}]
			{\includegraphics[width=0.45\textwidth, clip, keepaspectratio]{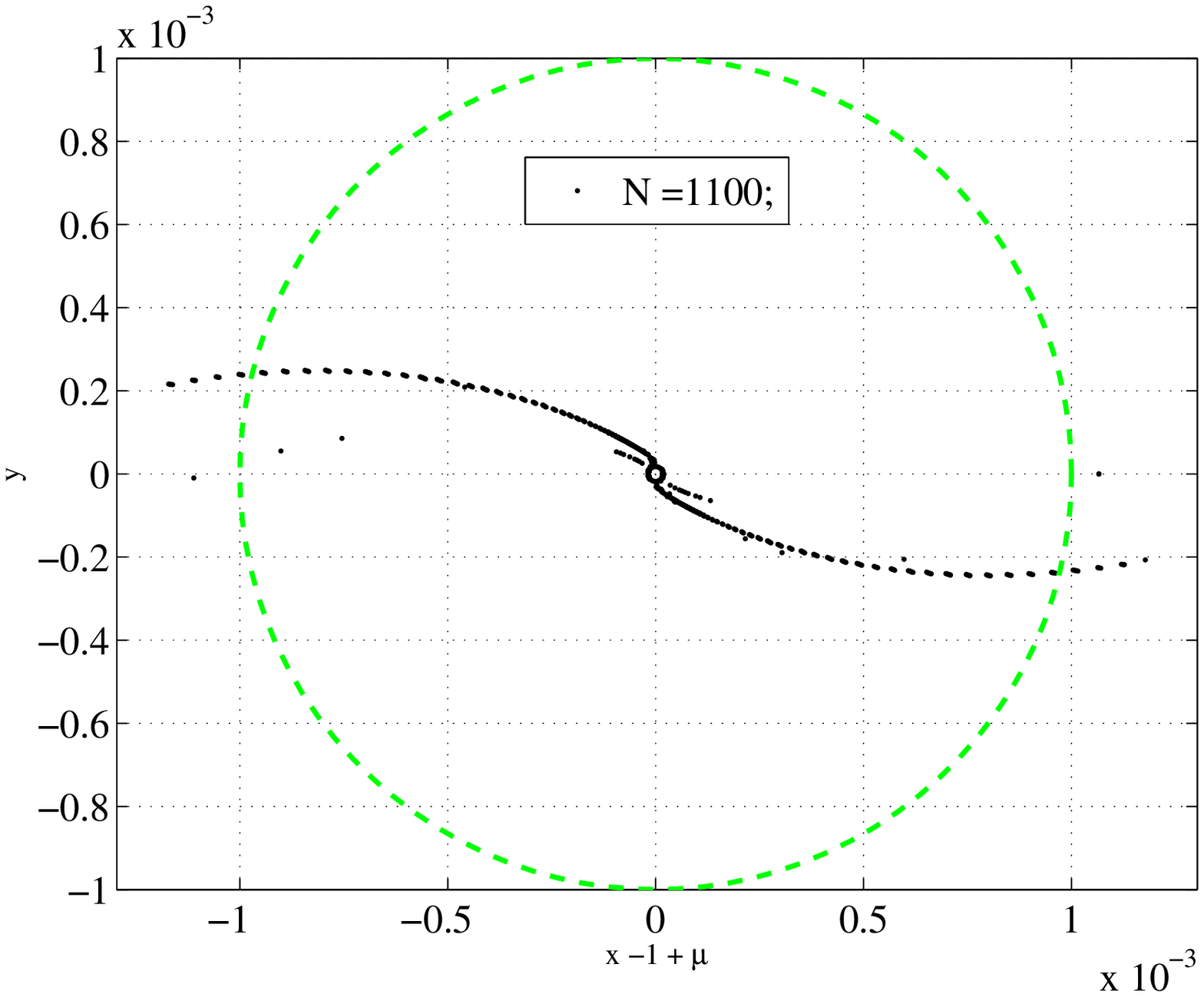}}
\caption{Sample stable sets $\W_{n}(e, f_0)$ for $f_0 = \pi/4$, $e = 0.99$, $n=1,2,3,6$ ($n=4,5$ not reported for brevity). $N$ is the number of stable initial conditions, whereas the green dashed circle represents the Levi-Civita regularizing disc. Figures corresponding to $n=1,2,3,6$ are read from left to right, top to bottom.}
\label{fig:StableSets}
\end{figure}

\begin{figure}[t!]
\centering
	\includegraphics[width=0.65\textwidth, clip, keepaspectratio]{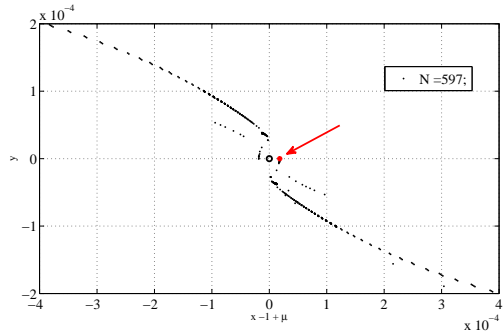}
	    \caption{Capture set $\C_{-1}^6(0.99,\pi/4)$.}
    \label{fig:CaptureSet}
\end{figure}

\begin{figure}[t!]
\centering
	\subfigure[Mars-centered rotating frame \label{fig:CapOrbit1-rot}]
		{\includegraphics[width=0.45\textwidth, clip, keepaspectratio]{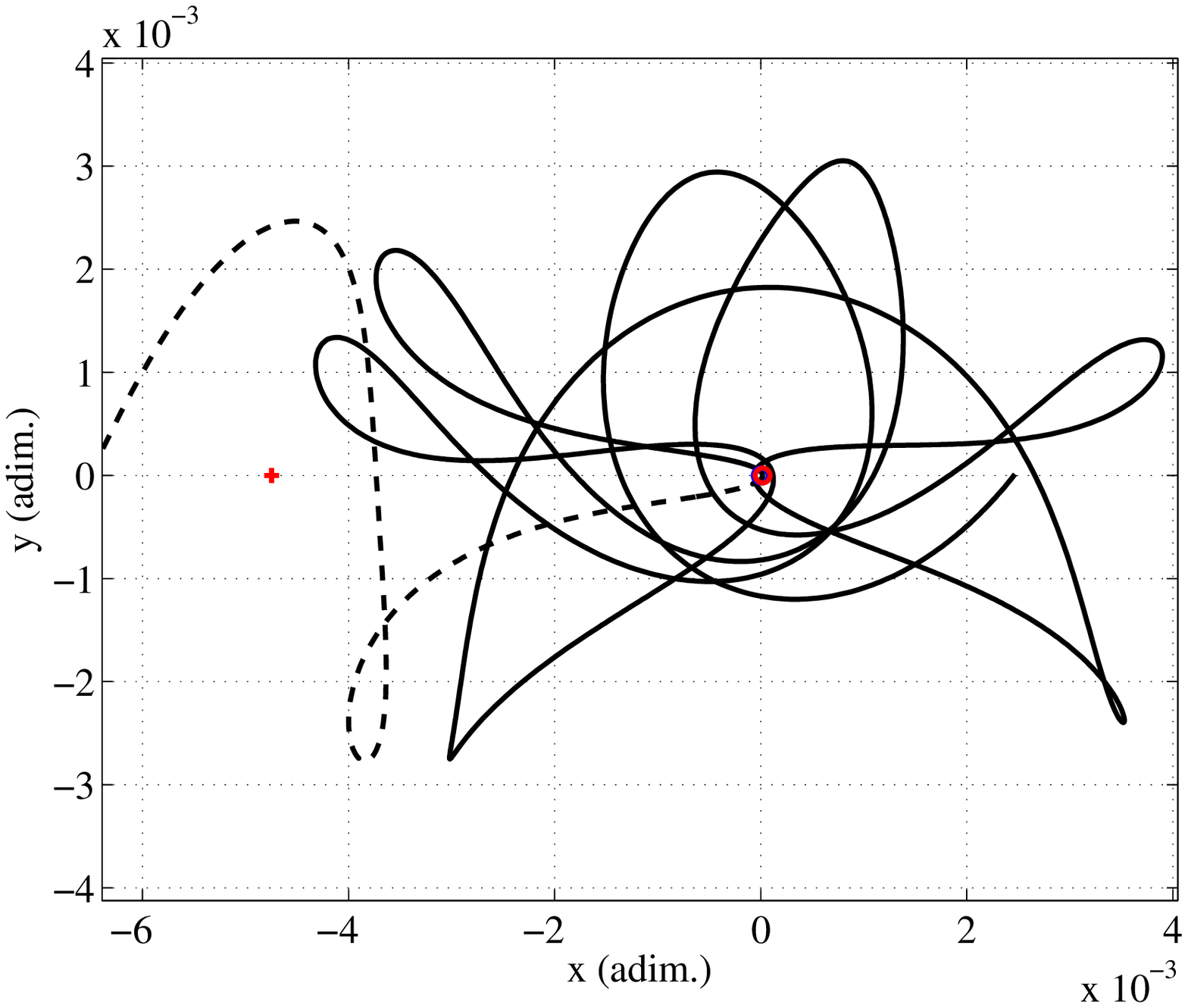}}
	\subfigure[Mars-centered inertial frame \label{fig:CapOrbit1-ine}]
		{\includegraphics[width=0.45\textwidth, clip, keepaspectratio]{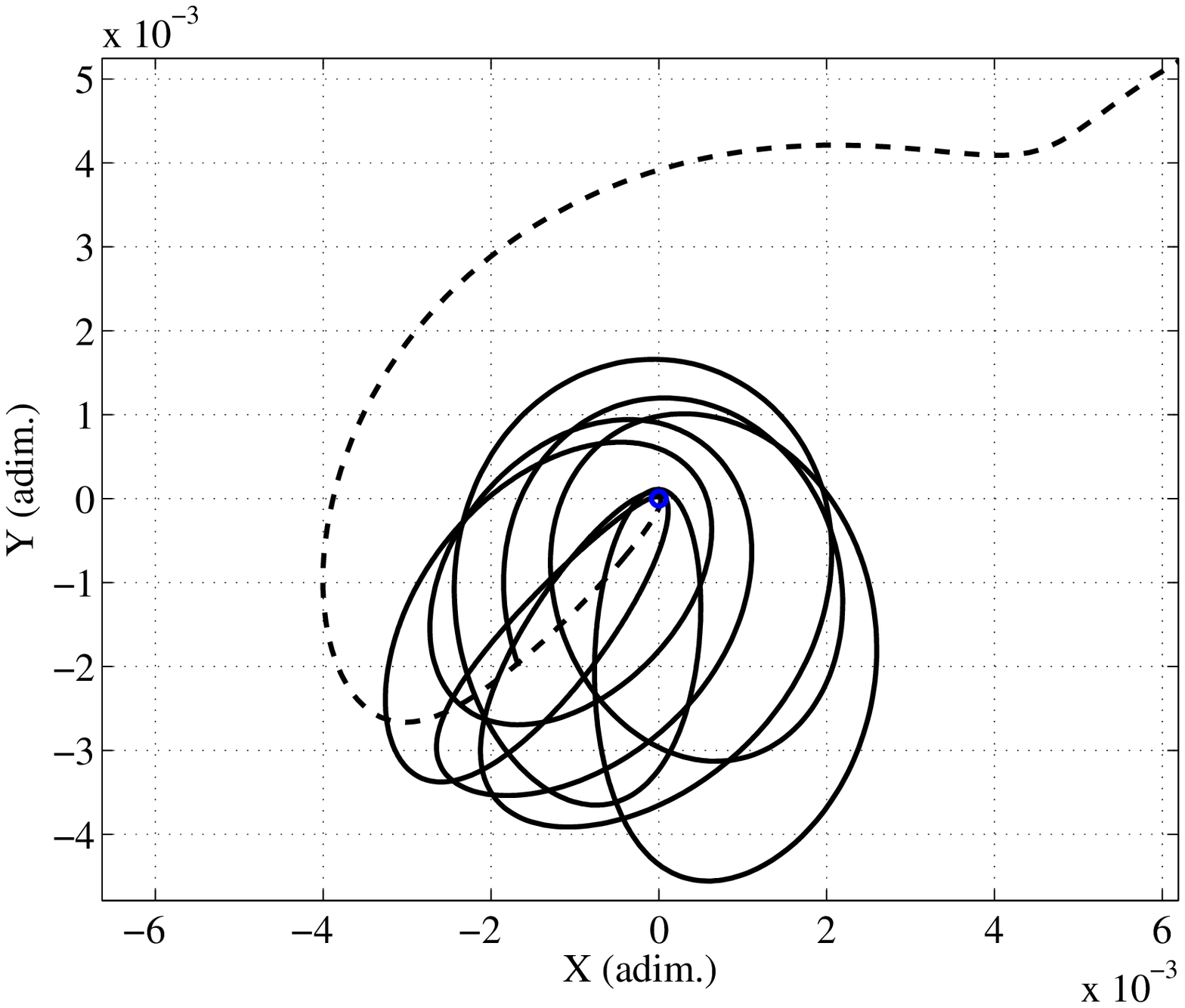}}
	\subfigure[Rotating frame, zoom out \label{fig:CapOrbit1-rot_zoomout}]
		{\includegraphics[width=0.45\textwidth, clip, keepaspectratio]{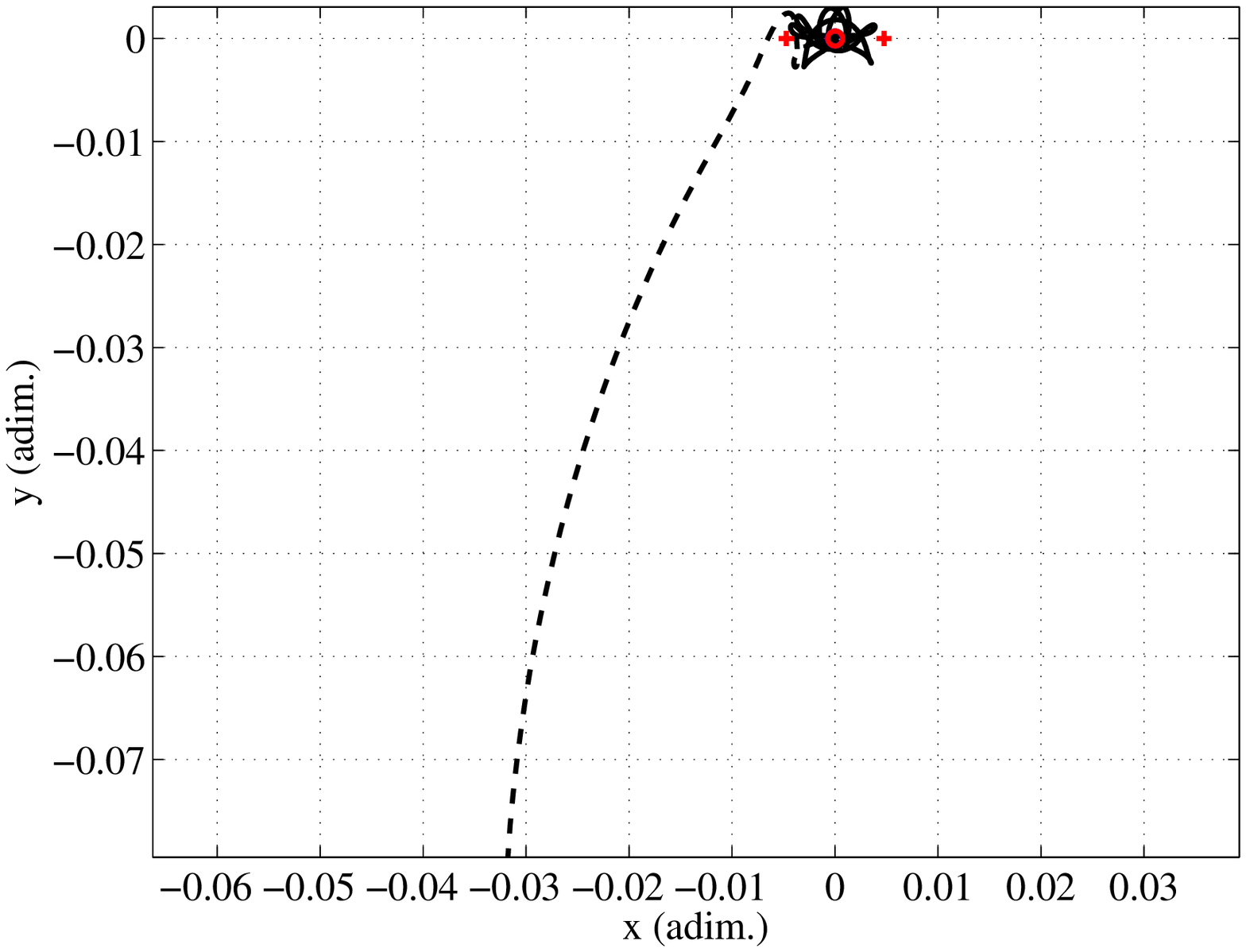}}
	\subfigure[Sun-centered inertial frame \label{fig:CapOrbit1-ine_sun}]
		{\includegraphics[width=0.45\textwidth, clip, keepaspectratio]{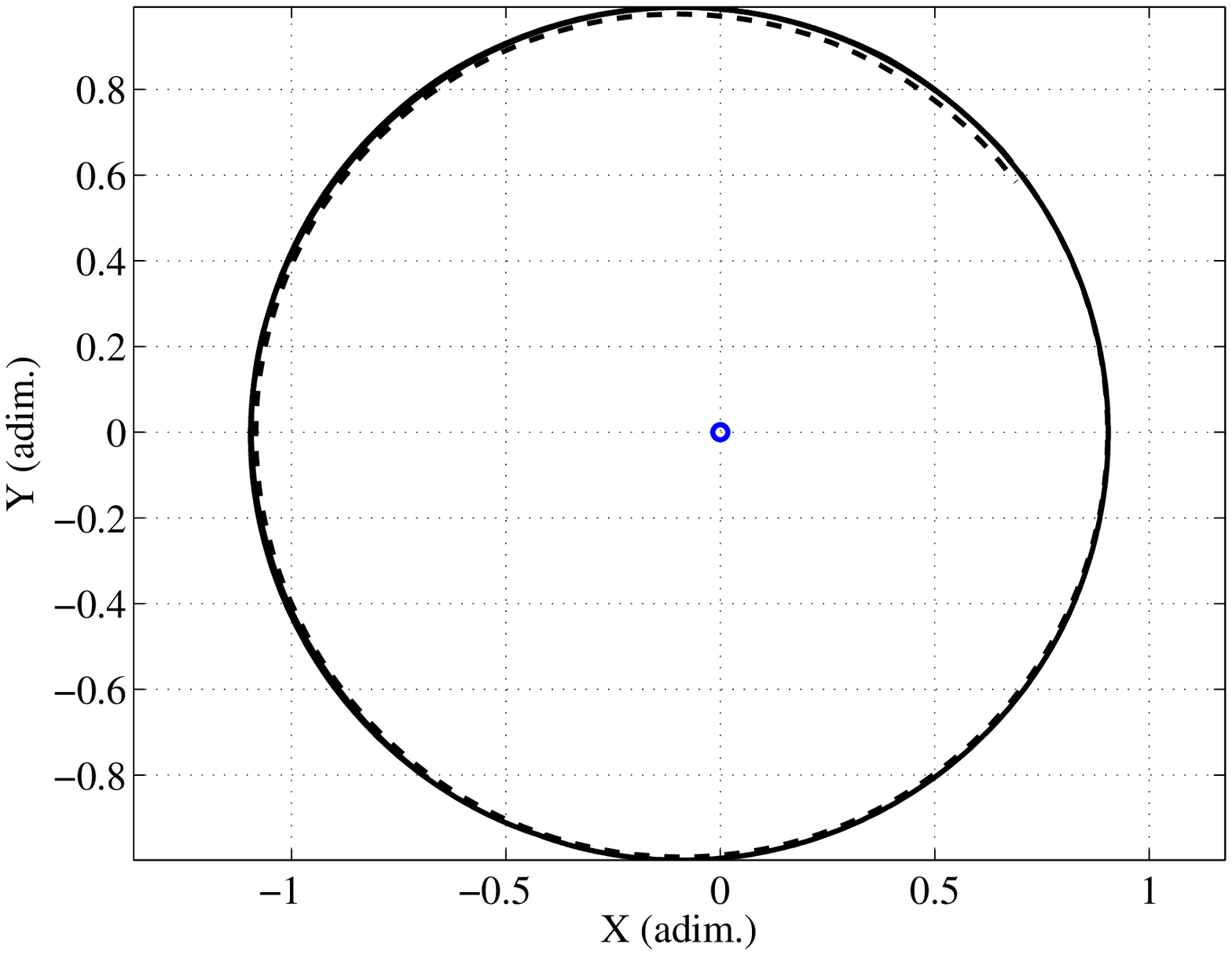}}
\caption{Capture orbit corresponding to a point in the set $\C_{-1}^6(0.99,\pi/4)$ (the point indicated in Figure \ref{fig:CaptureSet}) in rotating and inertial (Mars- and Sun-centered) coordinates.}
    \label{fig:CapOrbit1}
\end{figure}

Some results from \cite{hyeraci2010} are recalled. The stable set $\mathcal{W}_{n}(e, f_0)$ is shown in Figure \ref{fig:StableSets} for different $n$, and given values of $e,f_0$. To generate these plots, $N$ stable points are plotted. The capture set $\C_{-1}^6(0.99,\pi/4)$ associated to the set in Figure \ref{fig:StableSets} for $n=6$ is shown in Figure \ref{fig:CaptureSet}. Each point in $\C_{-1}^6(0.99,\pi/4)$ gives rise to an orbit that approaches Mars and performs \textit{at least} 6 revolutions around it. In Figure \ref{fig:CapOrbit1} the orbit generated by the point indicated in Figure \ref{fig:CaptureSet} is shown in several reference frames. If a spacecraft moved on this orbit, it would approach Mars on the dashed curve and it would remain temporarily trapped about it (solid line) without performing any maneuver. The trajectory represented by the dashed curve is a ballistic capture trajectory, or transfer, approaching the ballistic capture state that gives rise to capture orbits.

If needed, the spacecraft could then be placed into a more stable orbit within the time frame of the temporary capture, so avoiding the hazards associated to single-point injections, typical of hyperbolic approaches. From this example it is clear that this concept relies on a simple definition of stability and manipulation of the stable sets. The strength of the method lies in its simplicity, and its application in more complex modeling is straightforward. This is a significant departure from the use of invariant manifolds.

\subsection{Long Term Behavior of the Capture Orbits}

To design transfers that exploit the ballistic orbits contained in $\mathcal{C}_{-1}^n$, the long-term behavior of the capture orbits has to be analyzed. In particular, as the aim is to design transfers that target the capture orbits, their long-term behavior has to be evaluated. To do that, we have integrated the capture orbit in Figure \ref{fig:CapOrbit1} backward in time for a time span equal to 50 revolutions of Mars around the Sun; i.e., 34,345 days or equivalently about 94 years. Of course, this time span is not comparable to that of a practical case, but it is anyway useful to check the long-term behavior of the capture orbits within such time interval to infer features on its dynamics.

\begin{figure}
\centering
	\includegraphics[width=0.65\textwidth, clip, keepaspectratio]{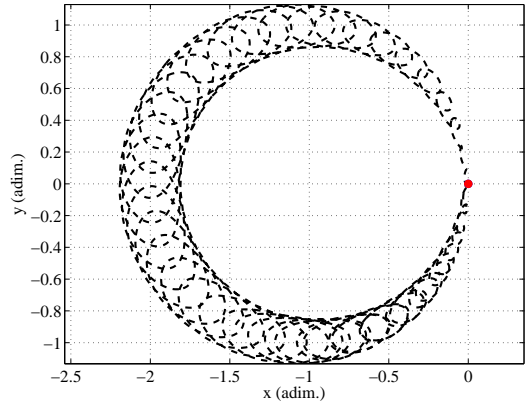}
	    \caption{Approaching portion of Figure \ref{fig:CapOrbit1} (dashed line) integrated backward for a time equal to 50 revolutions around Mars (Sun-Mars rotating frame).}
    \label{fig:LongTerm}
\end{figure}

As it can be seen from Figure \ref{fig:LongTerm}, the capture orbit gets close to Mars (red dot). This happens approximately 80 years backward in time from the ballistic capture occurrence. Although it approaches Mars, the capture orbit does not enter the Mars region, and therefore there is not a second ballistic capture. The most interesting behavior is that, although integrated backward for almost a century, the ballistic capture orbit does not substantially go far from the orbit of Mars. It is as if the phasing with Mars changes, but the third body is still trapped about Mars region.

\subsection{Constructing Ballistic Capture Transfers Starting Far From Mars}

Of particular interest in this paper is to find ballistic capture transfers that start far from Mars. (These results are new and not obtained in \cite{hyeraci2010}.) This is conveniently done by integrating the ballistic capture states in Figure \ref{fig:CaptureSet} and see where they go. We find that these trajectories, in backwards time, move far from Mars, but close to Mars orbit about the Sun. Their terminal point is the target for our transfers departing from the Earth.

For the sake of an example, consider the point indicated in Figure \ref{fig:CaptureSet}, which belongs to the set  $\C_{-1}^6(0.99,\pi/4)$. The forward and backward integrations are reported in Figure \ref{fig:CapOrbit1} and projected onto different reference frames. When integrated forward (solid line), the orbit performs 6 orbits about Mars in a totally ballistic fashion (i.e., no maneuvers accounted for). When integrated backward (dashed line), the orbit leaves Mars, by definition, but stays in a near ballistic capture state about Mars. The global ballistic capture trajectory obtained by the backwards integration of the ballistic capture trajectory near to Mars shown in Figure \ref{fig:CapOrbit1-ine} is shown in Figure \ref{fig:CapOrbit1-rot_zoomout} and then more globally in Figure \ref{fig:CapOrbit1-ine_sun}.

In the next section, we will pick locations along the dashed line, near to Mars orbit, where to start the global ballistic capture transfer, that leads to ballistic capture and to the resulting capture orbits.

\section{Interplanetary Transfer from Earth to Capture Points Far From Mars} \label{sec:4}

The purpose of this section is to describe the construction of the transfer from the Earth to Mars at the ballistic capture point $\bf x_c$.  We show the full ballistic capture transfer from the Earth to Mars obtained by linking this up with a ballistic capture transfer to that goes to the distance $r_p$ for ballistic capture. We describe the dynamics of the capture process, which is interesting.  This comprises Step 2 and part of Step 3 in Section \ref{sec:1}. In Section \ref{sec:5} comparison to Hohmann at $r=r_p$ is given, completing Step 3.

\medskip

A point, $\bf x_c$, is chosen near the orbit of Mars from which to begin a ballistic capture orbit that will go to ballistic capture to Mars at a periapsis distance $r_p$. We choose it in an arbitrary fashion, but to be beyond the SOI of Mars, so that the gravitational force of Mars there is negligibly small. This point is obtained by integrating a ballistic capture orbit from $r_p$ in backwards time so that it moves sufficiently far from Mars.   An example of this is seen in Figure \ref{fig:CapOrbit1-ine_sun} for the particular capture trajectory shown in the previous section, Section \ref{sec:3}. In that case we choose $\bf x_c$  about 1 million km from Mars. (see Figure \ref{fig:BalCapOrbits1}) When we consider different capture trajectories in this case with different properties, such as different values of $r_p$, then as the trajectory is integrated backwards for different $r_p$, the different trajectories will all have values of  $\bf x_c$ that lie very close to each other. So, for each of the different ballistic capture transfers for a given case, such as that shown in  Figure \ref{fig:CapOrbit1-ine_sun} we refer to as Case 1,  we will allow $\bf x_c$ to slightly vary.  

We will also generate another complete Earth to Mars ballistic capture transfer where $\bf x_c$ is much further from Mars, at a distance of about $23$ million km, that is shown in Figure \ref{fig:sol12-ine}. We refer to this as Case 2.  There are many possibilities for the choice of $\bf x_c$ but in this paper, we have chosen the two locations at $1$ and $23$ million km from Mars, respectively, for the sake of argument.

\subsection{Dynamics of Capture and Complete Transfer from Earth to Mars Ballistic Capture}

\begin{figure}[t!]
	\centering
		\hspace{-5mm}\ \subfigure[Transfer from Earth to ${\bf x_c}$ \label{fig:TransEMarsBalCap1}]
			{\includegraphics[width=0.5\textwidth, clip, keepaspectratio]{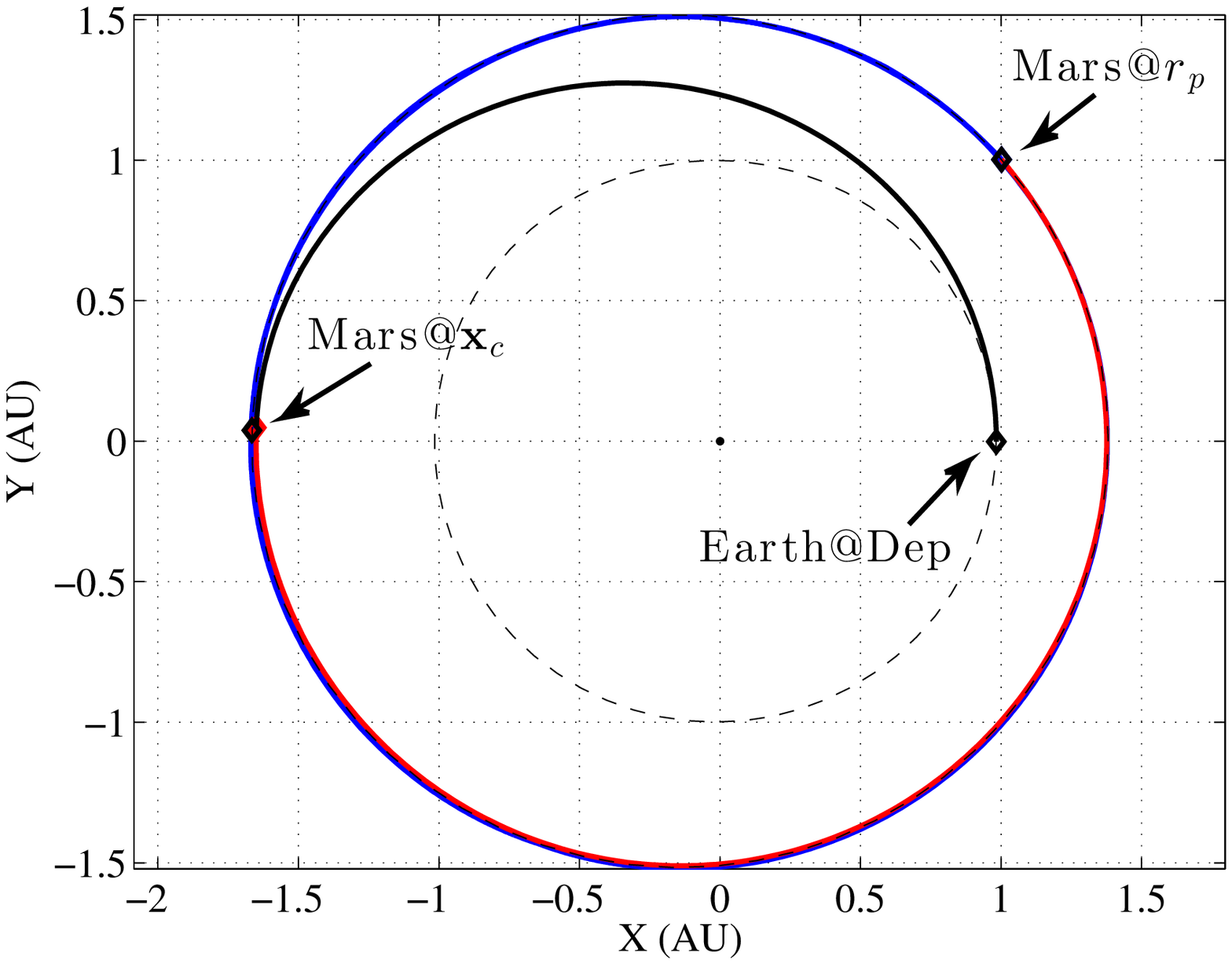}}
			\subfigure[Transfer from ${\bf x_c}$ to ballistic capture \label{fig:BalCapOrbits1}]
			{\includegraphics[width=0.5\textwidth, clip, keepaspectratio]{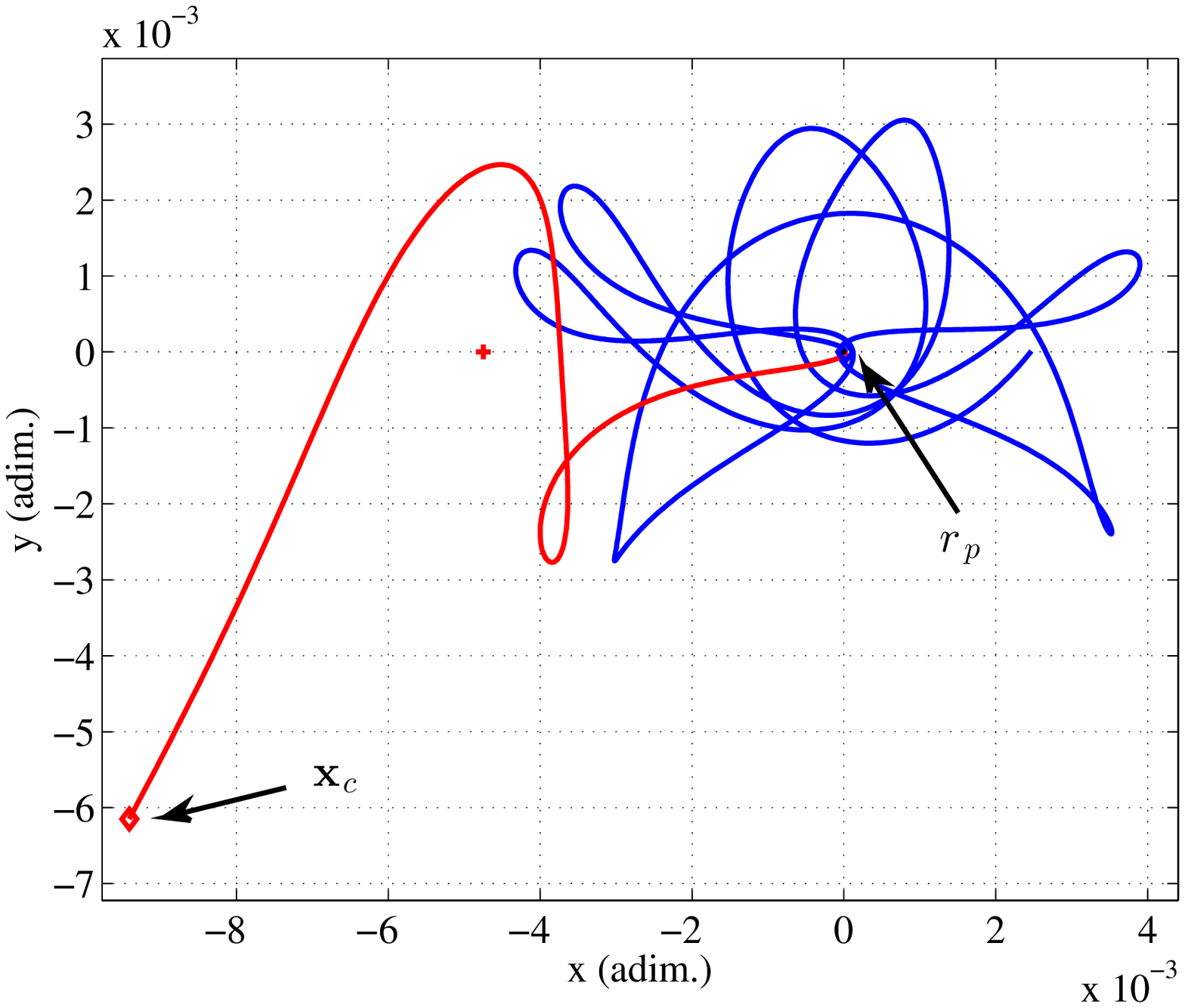}}
\caption{A sample solution constructed by using the orbit in Figure \ref{fig:CapOrbit1}. Left: Sun-centered frame (the black orbit is the orbit needed to target the capture point departing from the Earth; the red orbit is the capture orbit; the blue orbit is the post-capture orbit). Right: the capture orbit (red) and the post-capture orbit (blue) in the rotating Mars-centered frame.}
\label{fig:CompleteTransfer1}
\end{figure}

The interplanetary transfer together with the ballistic capture transfer comprise a ballistic capture transfer from the Earth to Mars.  An example of this is given in Figure \ref{fig:CompleteTransfer1} for Case 1.  The location of Mars when the spacecraft, $P$, arrives at $\bf x_c$ is indicated. As can be seen, Mars is initially behind $\bf x_c$ and about  $1$ million km away. However, Mars is moving slightly faster than $P$ as $P$ leaves $\bf x_c$ on the ballistic capture transfer to the distance $r_p$ from Mars. Approximately a year later, $P$ is overtaken by Mars, then $P$ catches up to Mars for ballistic capture at $r_p$ into a set of capture orbits moving at least 6 orbits about Mars within the stable set. The capture dynamics near Mars is illustrated in Figure \ref{fig:BalCapOrbits1} where the capture transfer remains below the Mars--Sun line, then slightly above the line then below where it is captured.  This approximately one year transit time of the ballistic capture transfer could be be significantly reduced if at $\bf x_c$ a tiny $\Delta V$ were applied to very slightly decrease the velocity of the spacecraft about the Sun. Then, Mars would catch up faster. This analysis is out of the scope of this paper and left for future study.

\begin{figure}[h!]
	\centering
		\subfigure[Inertial frame \label{fig:sol12-ine}]
			{\includegraphics[width=0.49\textwidth, clip, keepaspectratio]{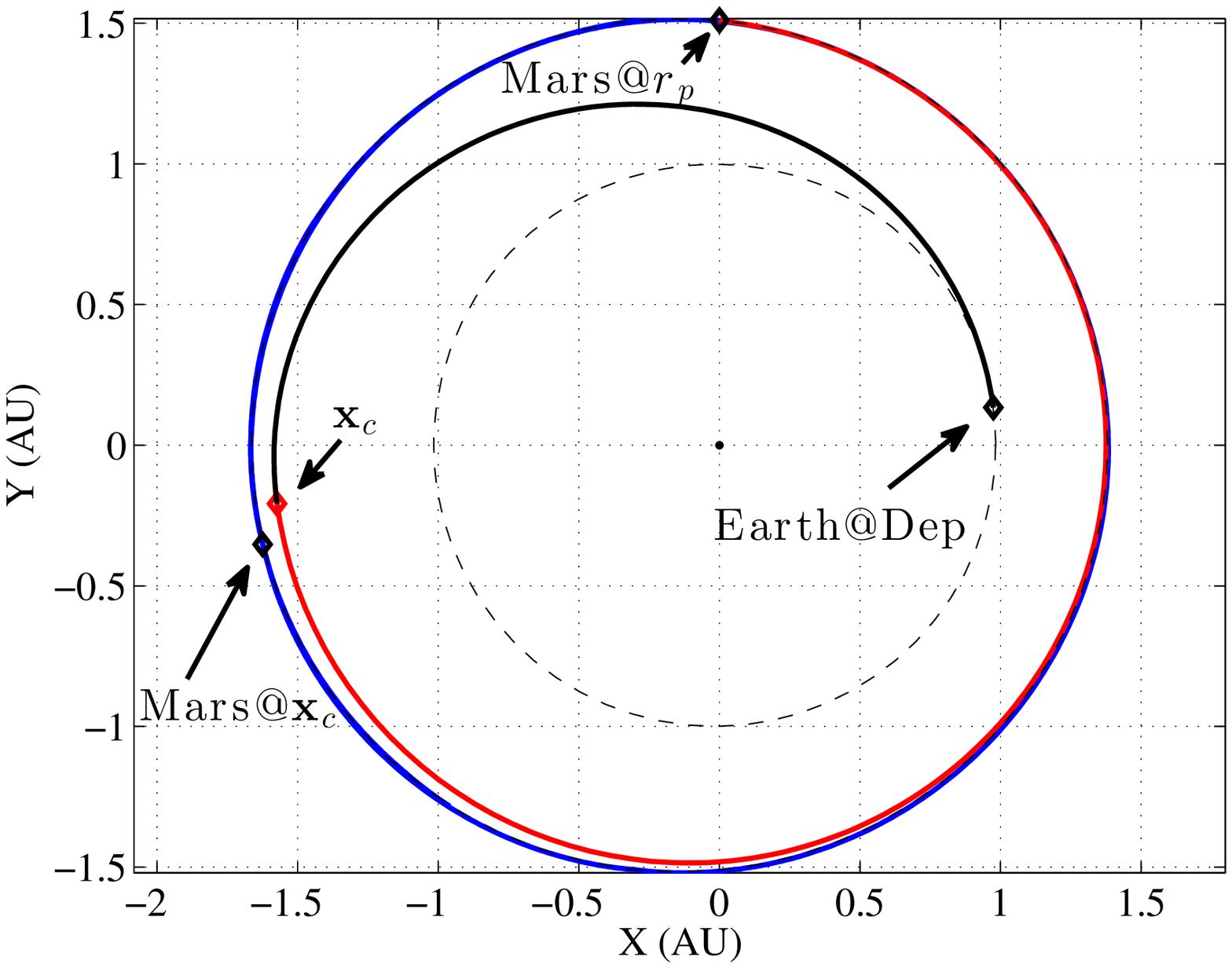}}
		\subfigure[Rotating frame \label{fig:sol12-rot}]
			{\includegraphics[width=0.49\textwidth, clip, keepaspectratio]{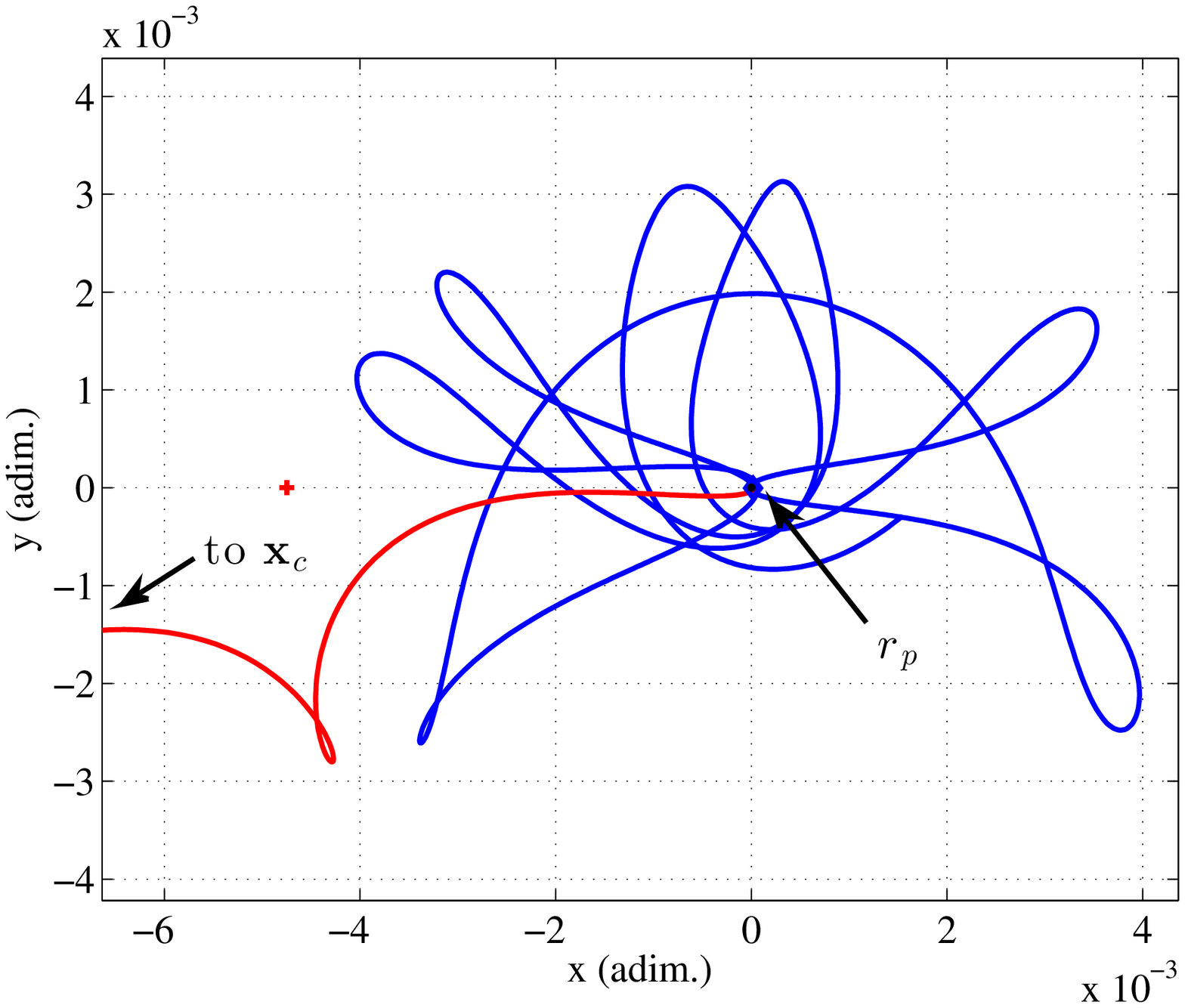}}
\caption{A sample solution obtained by targeting a point in $\C_{-1}^6(0.99,\pi/2)$. This solution is particularly interesting as it presents a quick backward escape: the target point $\bf x_c$ is $23\times10^6$ km far from Mars. Left: Sun-centered inertial frame. Right: rotating Mars-centered frame.}
\label{fig:sol12}
\end{figure}

Another example of a complete ballistic capture transfer from the Earth is shown in Figure \ref{fig:sol12} for Case 2.  Here, the dynamics of capture is different than in the previous case. When the spacecraft arrives at $\bf x_c$, Mars ahead of $\bf x_c$.  In this case, the spacecraft is initially moving faster than Mars. It eventually overtakes Mars and then is pulled back towards Mars for ballistic capture in about $1$ year.  

\subsection{Optimization of Transfers from Earth to Mars Ballistic Capture}

The transfers from Earth to Mars ballistic capture orbits are sought under the following assumptions. 1) The equations describing the ballistic capture dynamics are those of the planar, elliptic restricted three-body problem; 2) The whole transfer is planar, that is, the Earth and Mars are assumed to revolve in coplanar orbits; 3) A first maneuver, $\Delta V_1$, is performed to leave the Earth. This is computed by assuming the spacecraft as being already in heliocentric orbit at the Earth\rq{}s SOI; 4) A second maneuver, $\Delta V_c$, is performed to inject the spacecraft into the ballistic capture orbit; 5) In between the two maneuvers, the spacecraft moves in the heliocentric space far from both the Earth and Mars, and therefore the dynamics is that of the two-body problem \cite{topputo2005b}.

The \textit{parameters} of the optimization (to be picked and held fixed) are:
\begin{itemize}
\item The Capture set. The stable sets computed keep fixed eccentricity. Moreover, when the capture sets are defined from the stable sets, the stability number has to be decided. Therefore, selecting the capture sets means fixing 1) the osculating eccentricity of the first post-capture orbit; 2) the stability number; i.e., the minimum number of natural revolutions around Mars.
\item The initial capture orbit within the set. For example, this is equivalent to specifying the radial and angular position for each of the black dots in Figure \ref{fig:CaptureSet}, and choosing one of these. This selection yields an integer number, $N$.
\end{itemize}

The \textit{variables} of the optimization problem are
\begin{itemize}
\item The time of the backward integration. This time is needed to define $\bf x_c$ (the target point) by starting from $r_p$ and performing a backward integration.
\item The time of flight from the Earth to the target point $\bf x_c$. This is needed to solve the Lambert problem once the position of the Earth is known.
\item A phase angle to specify the position of the Earth on its orbit.
\end{itemize}

The \textit{objective function} is the cost of the second maneuver, $\Delta V_c$.  It is assumed that the first maneuver, $\Delta V_1$, can be always achieved, whatever it costs. Moreover, it is expected that the cost for $\Delta V_1$ is equivalent to that of a standard Hohmann transfer as the target point is from an angular perspective, not too far from Mars.

\section{Comparison of Ballistic Capture Transfer to Hohmann} \label{sec:5}

The parameters for the reference Hohmann transfers from Earth SOI to Mars SOI are listed in Table \ref{tab:Hcases} in Appendix 2; these figures correspond to geometries where four different bitangential transfers are possible. The hyperbolic excess velocity at Mars SOI for these bitangential transfers are listed in Table \ref{tab:Vinf}. These will be taken as reference solutions to compare the ballistic capture transfers derived in this paper. These four reference solutions represent a lower bound for all possible patched-conics transfers: when the transfer orbit is not tangential to Mars orbit, the hyperbolic excess velocity increases.

\begin{table}[h!]
\caption{Hyperbolic excess velocities at Mars for the four bitangential transfers.}
\medskip

\label{tab:Vinf} 
\centering
\begin{tabular}{cc}
\hline
Case	& $V_\infty$ (km/s) \\
\hline
H1		& 3.388	 \\
H2		& 2.090	 \\
H3		& 3.163	 \\
H4		& 1.881	 \\
\hline
\end{tabular}
\end{table}

When approaching Mars in hyperbolic state with excess velocity $V_\infty$ at Mars SOI, the cost to inject into an elliptic orbit with fixed eccentricity $e$ and periapsis radius $r_p$ is straight forward to compute as,
\begin{equation} \label{eq:Hcost}
	\Delta V_2 = \sqrt{V_\infty^2 + \dfrac{2 \mu_M}{r_p}} - \sqrt{\dfrac{\mu_M (1+e)}{r_p}}
\end{equation}
where $\mu_M$ is the gravitational parameters of Mars (see Table \ref{tab:constants}, Appendix 2). This formula is used to compute the $\Delta V_2$ for different values of $r_p$. 

\bigskip

It is important to note that the main goal of this paper is to study the performance of the ballistic capture transfers from the Earth to Mars from the perspective of the capture $\Delta v$ as compared to Hohmann transfers, when going to specific periapsis radii, $r_p$.  This is done irrespective of $\Delta V_1$. However, in the case we are doing a detailed analysis, $\bf x_c$ is $1$ million km from Mars, and because of this, the value of $\Delta V_1$ for both Hohmann and Ballistic capture transfers are approximately the same.  This should also be the case in the other complete transfer computed where $\bf x_c$ is $23$ million km from Mars. Thus, in these cases, studying the capture $\Delta v$ performance is equivalent to the total $\Delta v$ performance.  However, this need not be the case if $\bf x_c$ is at a distance such as $200$ million km from Mars. The choice of such large distances for $\bf x_c$ are not considered in this paper and are for future study.

\bigskip

\begin{figure}
	\centering
	\hspace{-6mm}
		\subfigure[$f_0 = 0$ \label{fig:BCvsH_e99_f000}]
			{\includegraphics[width=0.36\textwidth, clip, keepaspectratio]{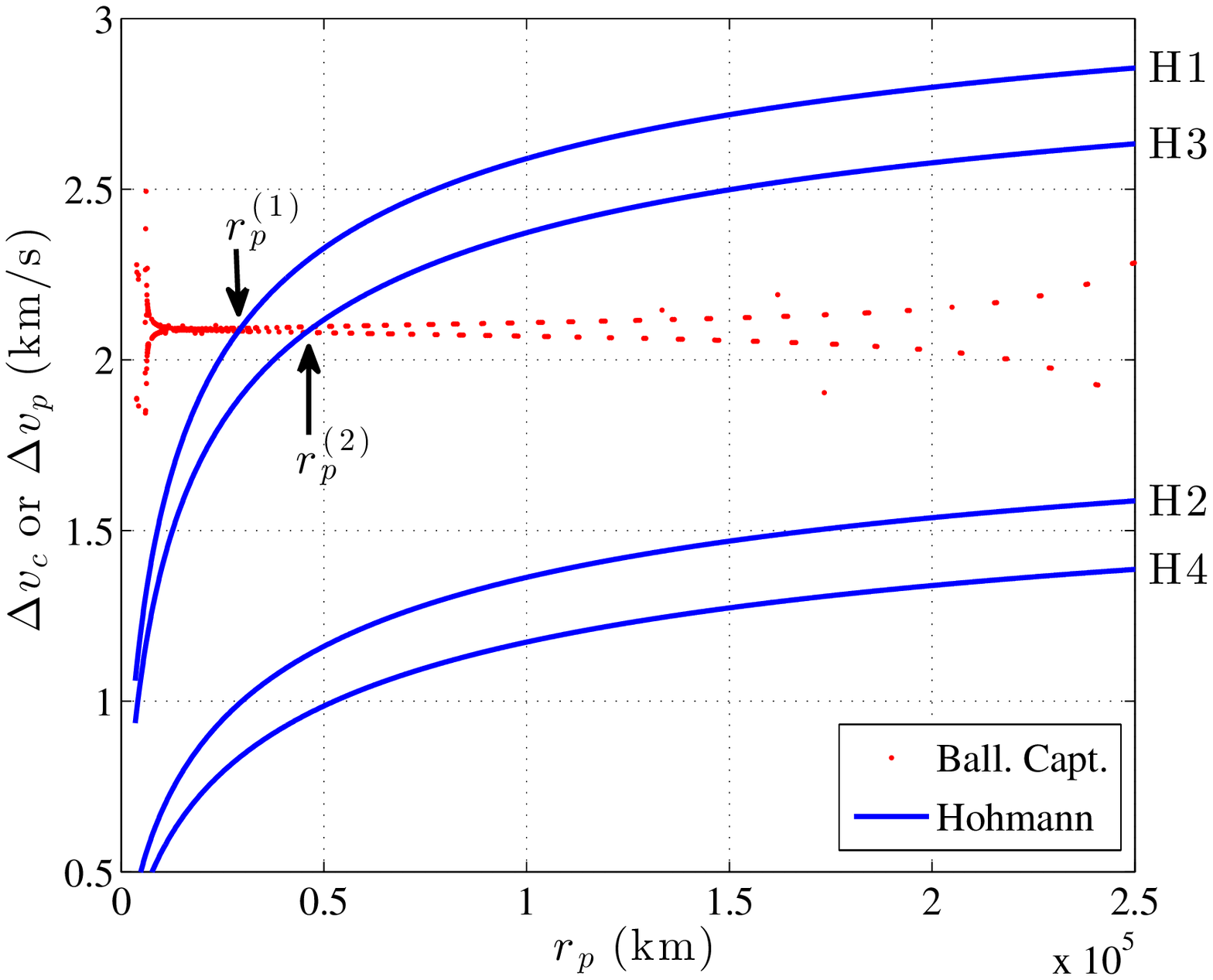}}
	\hspace{-6mm}
		\subfigure[$f_0 = \pi/4$ \label{fig:BCvsH_e99_f025}]
			{\includegraphics[width=0.36\textwidth, clip, keepaspectratio]{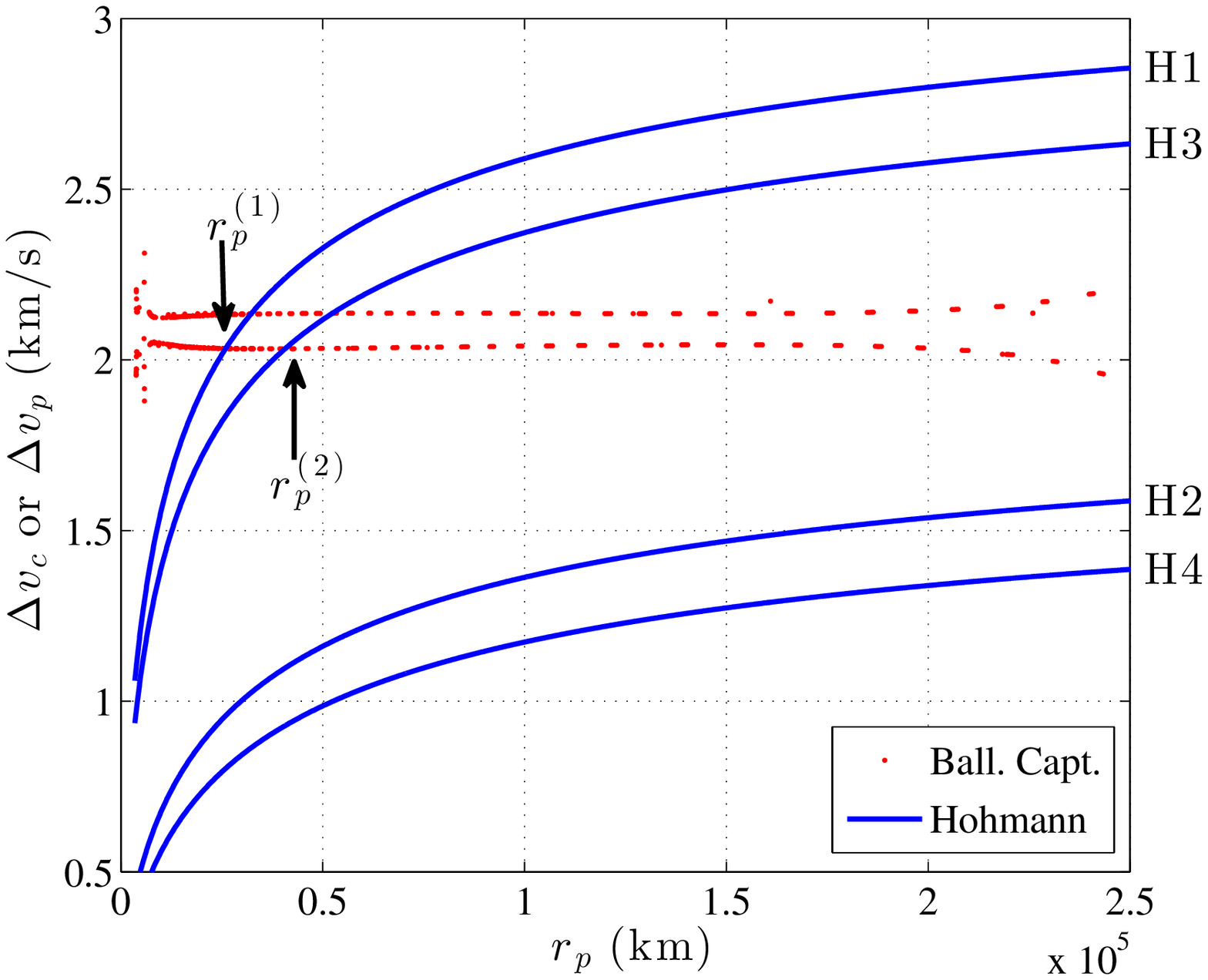}}
	\hspace{-6mm}	
		\subfigure[$f_0 = \pi/2$ \label{fig:BCvsH_e99_f050}]
			{\includegraphics[width=0.36\textwidth, clip, keepaspectratio]{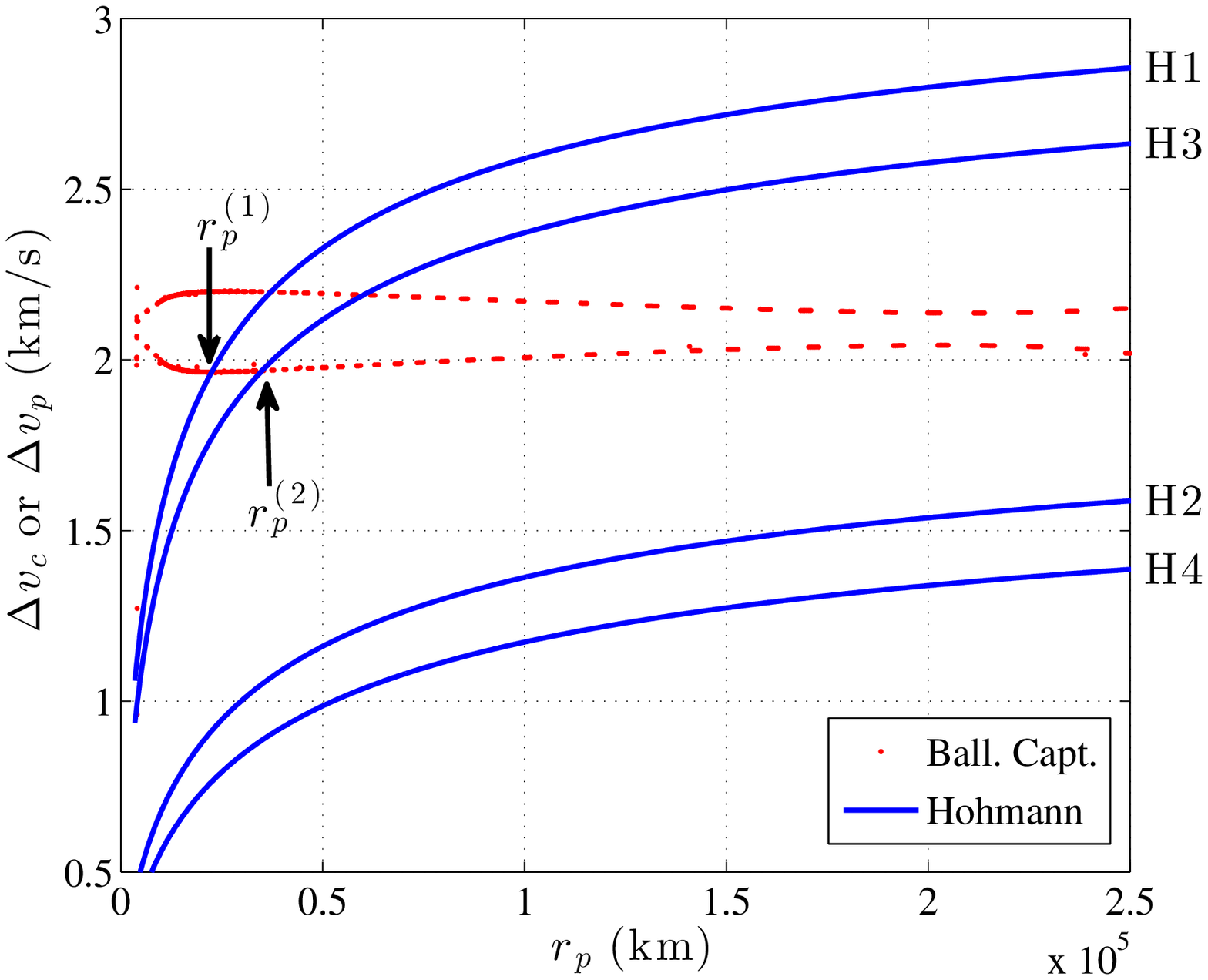}}
\caption{Comparison of Hohmann bitangential transfers and ballistic capture transfers originated by the capture sets $\mathcal{C}(e, f_0)$, $e=0.99$, $f_0 = 0, \pi/4, \pi/2$.}
\label{fig:BCvsH_e99}
\end{figure}

An assessment of the ballistic capture transfers whose $\bf x_c$ states are originated by the sets $\mathcal{C}(e, f_0)$, with $e=0.99$ and $f_0 = 0, \pi/4, \pi/2$, has been made. The results are summarized in Figure \ref{fig:BCvsH_e99}. In these figures, the red dots represent the $\Delta V_c$  cost of the ballistic capture solutions from the two cases, whereas the blue curves are the functions $\Delta V_2(r_p)$ computed from \eqref{eq:Hcost} associated to the four bitangential Hohmann transfers in Table \ref{tab:Hcases}. From inspection of Figure \ref{fig:BCvsH_e99} it can be seen that the ballistic capture transfers are more expensive then all of the Hohmann transfers for low altitudes. Nevertheless, when $r_p$ increases, the ballistic capture transfer perform better than H1 and H3. This occurs at periapsis radii $r_p^{(1)}$ and $r_p^{(2)}$, respectively, whose values are reported in Table \ref{tab:rp_change} along with the values for which $\Delta V_c \simeq \Delta V_2$. For periapsis radii above $r_p^{(1)}$ or $r_p^{(2)}$, the savings increase for increasing $r_p$. In the cases of H2, H4, the ballistic capture transfers do not perform as well as the Hohmann transfers for any value of $r_p$. 

\begin{table}[h!]
\caption{Periapsis radii for which the ballistic capture transfers become more convenient than the Hohmann transfers for different $f_0$.}
\medskip

\label{tab:rp_change} 
\centering
\begin{tabular}{rccc}
\hline
$f_0$	& $r_P^{(1)}$ (km)	& $r_P^{(2)}$ (km) 	& $\Delta V_c$ (km/s)\\
\hline
0		& $29\times10^3$	& $46\times10^3$	& 2.09	 			\\
$\pi/4$	& $26\times10^3$	& $40\times10^3$	& 2.03				\\
$\pi/2$	& $22\times10^3$	& $34\times10^3$	& 1.96				\\
\hline
\end{tabular}
\end{table}

A number of observations arise from the assessment performed. These are briefly given below.
\begin{itemize}
\item The cost for the ballistic capture transfers is approximately constant regardless of the periapsis radius $r_p$. This is a great departure from Hohmann transfers where the cost increases for increasing $r_p$.
\item The red dots in Figure \ref{fig:BCvsH_e99} are organized into two different sets that correspond to the two branches of the capture sets, see Figure \ref{fig:CaptureSet}.
\end{itemize}
\medskip

The results from Figure \ref{fig:BCvsH_e99_f025} are summarized in Table \ref{tab:cases}. 

\begin{table}[h!]
\caption{Comparison between ballistic capture transfers and Hohmann transfers for the points in Figure \ref{fig:BCvsH_e99_f025}. The saving, $S$, is computed as $S=(\Delta V_c - \Delta V_2)/\Delta V_c$, where the $\Delta V_2$ associated to the H3 case is considered. $S$ is a measure of the efficiency of the ballistic capture transfers. $\Delta t_{c\to p}$ is the time-of-flight needed to go from $\bf {x}_c$ to $r_p$.}
\medskip

\label{tab:cases} 
\centering
\begin{tabular}{rccccc}
\hline
Point	& $r_P$ (km)	& $\Delta V_c$ (km) & $\Delta V_2$ (km/s)	& $S$ (\%) & $\Delta t_{c\to p}$ (days)\\
\hline
(A)		& $49896$		& $2.033$			& 2.116					& -4.0\%	& 434	 					\\
(B)		& $73896$		& $2.036$			& 2.267					& -11.3\%	& 433						\\
(C)		& $91897$		& $2.039$			& 2.344					& -14.9\%	& 432						\\
(D)		&$113897$		& $2.041$			& 2.414					& -18.2\%	& 431						\\
\hline
\end{tabular}
\end{table}

From this table it can be seen that the time for the spacecraft to go from $\bf x_c$ to $r_p$ is on the order of a year.  This time should be able to be decreased by very slightly adjusting $\Delta V_c$ so that the distance between the spacecraft and Mars decreases more rapidly. (The location of the points, A, B, C, D in Figure \ref{fig:BCvsH_e99_f025} only span a limited range of $r_p$ values. The percentage savings, S, would substantially increased for higher values of $r_p$.) 
\medskip

It is remarked that in the cases considered for $e=.99$, as the capture orbits cycle about Mars with high periapsis values, they will have apoapsis values beyond the SOI of Mars.  Since the SOI is purely a geometric definition and not based on actual dynamics, these ellipses are well defined outside of the SOI. The fact they exist in the elliptic restricted problem demonstrates this. 
\bigskip

In summary, we have the following,
\medskip

\noindent
{\em Result A} \hspace{.1in}  The ballistic capture transfers use less $\Delta V$ for the capture process than a Hohmann transfer for altitudes above $r_p^{(1)}, r_p^{(2)}$ in the cases for H1, H3 in the examples given, where 
\begin{equation}
\Delta V_c < \Delta V_2.
\label{eq:DVComparison}
\end{equation}
The percentage savings in these cases can be on the order of $25\%$ when $r_p$ is $200,000$ km. 
\medskip

\subsection{Transfer to Low Values of $r_p$, Launch Period Flexibility}

The fact that one can have $\bf x_c$ far from Mars has an implication on the launch period from the Earth to get to Mars.  For the case of a Hohmann transfer, there is a small launch period of a few days that must be satisfied when the Mars and Earth line up. This is because a point, i.e. Mars, has to be directly targeted. If this is missed for any reason, a large penalty in cost may occur since launch may not be possible.  This problem would be alleviated if the launch period could be extended.  By targeting to $\bf x_c$ rather than to Mars, it is not necessary to wait every two years, but rather, depending on how far $\bf x_c$ is separated from Mars, the time of launch could be extended significantly.  This is because an orbit is being targeted, rather than a single point in the space.

This launch period flexibility has another implication. As determined in this paper, the Hohmann transfer is cases H1, H3 uses more capture $\Delta v$ than a ballistic capture transfer when $r_p > r_p^{(1)}, r_p^{(2)}$.  Since the capture $\Delta v$ used by the ballistic capture transfer and the Hohmann tansfer is the same when $r_p = r_p^{(1)}, r_p^{(2)}$, then the penalty, or excess, $\Delta v$ that a ballistic capture uses relative to a Hohmann transfer when transferring to a lower altitude can be estimated by just calculating the $\Delta v$'s to go from  a ballistic capture state at $r_p=r_p^{(1)}, r_p^{(2)}$ to a desired altitude lower than these, say to an altitude of $100$ km, where $r_p=r_p^* = 100 + r_M$, $r_M$ = radius of Mars.  

For example, lets consider the case where we transfer from $r_p = 40,000$ km to $r_p^*$.  To do this, it is calculated that the spacecraft must increase velocity by .196 km/s at $r_p$ and decrease velocity by  .192 km/s at $r_p*$.  This yields a total value of .380 km/s.  This number may be small enough to justify a ballistic capture transfer instead of a Hohmann transfer if it was decided that the flexibility of launch period was sufficiently important. 

\section{Summary, Applications and Future Work} \label{sec:6}

The capture $\Delta V$ savings offered by the ballistic capture transfer from the Earth to Mars is substantial when transferring to higher altitudes in certain situations. This may translate into considerable mass fraction savings for a spacecraft arriving at Mars, thereby allowing more payload to be placed into orbit or on the surface of Mars, over traditional transfers to Mars, which would be something interesting to study.  Although the Hohmann transfer provides lower capture $\Delta v$ performance in certain situations, in other cases it doesn\rq{}t, and in these the ballistic capture transfer offers a new approach. 

It isn\rq{}t the capture performance that is the only interesting feature.  The more interesting feature is that by targeting to points near Mars orbit to start a ballistic capture transfer, the target space opens considerably from that of a Hohmann transfer which must transfer directly to Mars. By transferring from the Earth to points far from Mars, the time of launch from the Earth opens up and is much more flexible. This flexibility of launch period offers a new possibility for Mars missions.    Also the methodology of first arriving far from Mars offers a new way to send spacecraft to Mars that may be beneficial from an operational point of view.  This launch flexibility and new operational framework offer new topics to study in more depth.

Another advantage of using the ballistic capture option is the benign nature of the capture process as compared to the Hohmann transfer.  The capture $\Delta v$ is done far from Mars and can be done in a gradual safe manner. Also, when the spacecraft arrives to Mars periapsis to go into orbit on the cycling ellipses, no $\Delta v $ is required. By comparison, the capture process for a Hohmann transfer needs to be done very quickly or the spacecraft is lost.  An example of this was with the Mars Observer mission.  In case low altitude orbits are desired, a number of injection opportunities arise during the multiple periapsis passages on the cycling ellipses. This is safer from an operational point of view to achieve low orbit, although only slightly more $\Delta v$ is used. 

Although the time of flight is longer as compared with a Hohmann transfer, this is only due to the choice of $\bf x_c$. By performing a minor adjustment to $\Delta V_c$, the time of flight to Mars should be able to be reduced, which an interesting topic to study for future work.

This new class of transfers to Mars offers new mission possibilities for Mars missions.

\section*{Acknowledgements}

We would like to thank the Boeing Space Exploration Division for sponsoring this work, and, in particular, we would like to thank Kevin Post and Michael Raftery.

\bibliography{bibliography}

\section*{Appendix 1}

\medskip

\noindent
{\em Summary of Precise Definitions of Stable Sets and Weak Stability Boundary}
\medskip

\begin{figure} 
\centering
	\includegraphics[width=0.6\textwidth, clip, keepaspectratio]{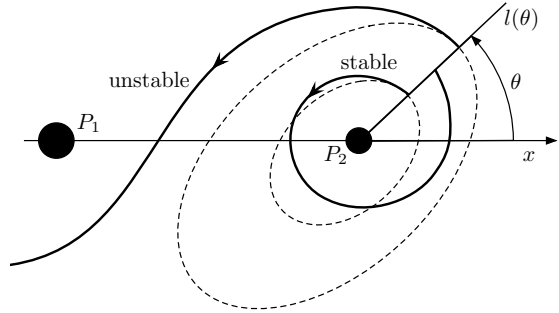}
	    \caption{Stable and unstable motions}
    \label{fig:StableUnstable}
\end{figure}

\noindent Trajectories of $P$ satisfying the following conditions are studied (see \cite{hyeraci2010,topputo2009a,garcia2007}).
\begin{itemize}
\item [(i)] The initial position of $P$ is on a radial segment $l(\theta)$ departing from $P_2$ and making an angle $\theta$ with the $P_1$--$P_2$ line, relative to the rotating system.  The trajectory is assumed to start at the periapsis of an osculating ellipse around $P_2$, whose semi-major axis lies on $l(\theta)$ and whose eccentricity $e$ is held fixed along $l(\theta)$.
\item [(ii)] In the $P_2$-centered inertial frame, the initial velocity of the trajectory is perpendicular to $l(\theta)$, and the Kepler energy, $H_2$, of $P$ relative to $P_2$ is negative; i.e., $H_2<0$ (ellipse periapsis condition). The motion, for fixed values of $e_p$, $f_0$, $\theta$, and $e$ depends on the initial distance $r$ only.
\item [(iii)] The motion is said to be $n$-stable if the infinitesimal mass $P$ leaves $l(\theta)$, makes $n$ complete revolutions about $P_2$, $n \ge 1$, and returns to $l(\theta)$ on a point with negative Kepler energy with respect to $P_2$, \textit{without} making a complete revolution around $P_1$ along this trajectory. The motion is otherwise said to be $n$-unstable (see Figure \ref{fig:StableUnstable}).
\end{itemize}

The set of $n$-stable points on $l(\theta)$ is a countable union of open intervals
\begin{equation} \label{eq:Wne-theta}
	\mathcal{W}_n(\theta, e, f_0) = \bigcup_{k \geq 1} (r^*_{2k-1}, r^*_{2k}),
\end{equation}
with $r_1^*=0$. The points of type $r^*$ (the endpoints of the intervals above, except for $r^*_1$) are $n$-unstable. Thus, for fixed pairs $(e, f_0)$, the collection of $n$-stable points is
\begin{equation} \label{eq:Wne}
	\mathcal{W}_n(e, f_0) = \bigcup_{\theta \in [0,2\pi]} \mathcal{W}_n (\theta, e, f_0).
\end{equation}

The weak stability boundary of order $n$, denoted by $\partial \mathcal{W}_n$, is the locus of all points $r^*(\theta, e, f_0)$ along the radial segment $l(\theta)$ for which there is a change of stability of the trajectory; i.e., $r^*(\theta, e, f_0)$ is one of the endpoints of an interval $(r^*_{2k-1}, r^*_{2k})$ characterized by the fact that, for all $r \in (r^*_{2k-1}, r^*_{2k})$, the motion is $n$-stable, and there  exist $\tilde r \not \in (r^*_{2k-1}, r^*_{2k})$, arbitrarily close to either $r^*_{2k-1}$ or $r^*_{2k}$ for which the motion is $n$-unstable. Thus,
\begin{displaymath}
	\partial \mathcal{W}_n(e, f_0) = \{r^*(\theta, e, f_0)\, |\, \theta \in [0,2\pi] \}.
\end{displaymath}

\section*{Appendix 2}

\medskip

\noindent
{\em Computation of  reference Hohmann transfers}
\medskip

\noindent The physical constants used in this work are listed in Table \ref{tab:constants}. As both the Earth and Mars are assumed as moving on elliptical orbits, there are four cases in which a bitangential transfer is possible, depending on their relative geometry. These are reported in Table \ref{tab:Hcases}, where `@P' and `@A' mean `at perihelium' and `at aphelium', respectively. In Table \ref{tab:Hcases}, $\Delta V_1$ is the maneuver needed to leave the Earth orbit, whereas $\Delta V_{2, \infty}$ is the maneuver needed to acquire the orbit of Mars; these two impulses are calculated by considering the spacecraft already in heliocentric orbit, and therefore $\Delta V_1$, $\Delta V_{2,\infty}$ are equivalent to the escape, incoming hyperbolic velocities, $V_{\infty}$, at Earth, Mars, respectively. $\Delta V$ and $\Delta t$ are the total cost and flight time, respectively.  The use of the notation, $\Delta V_{2,\infty}$ is to distinguish from the use of $\Delta V_2$ used in Section \ref{sec:5} for the actual $\Delta v$ used by the Hohmann transfer at the distance $r_p$.  

From the figures in Table \ref{tab:Hcases} it can be inferred that although the total cost presents minor variations among the four cases, the costs for the two maneuvers change considerably. This is important in this work where a quantitative comparison has to be made. That is, by arbitrary picking one of the four bitangential solutions as reference we can have different outcomes on the performance of the ballistic capture orbits presented in this paper.  Because there is a substantial variation, an averaging does not yield useful results, and therefore, each case is considered.

\begin{table}[h!]
\caption{Physical constants used in this work.}
\label{tab:constants} 
\centering
\begin{tabular}{rlll}
\hline
Symbol		& Value						& Units				& Meaning \\
\hline
$\mu_S$		& $1.32712\times10^{11}$	& ${\rm km^3/s^2}$	& Gravitational parameter of the Sun   \\
AU			& $149597870.66$			& ${\rm km}$		& Astronomical unit \\
\hline
$\mu_E$		& $3.98600\times10^{5}$		& ${\rm km^3/s^2}$	& Gravitational parameter of the Earth \\
$a_E$		& $1.000000230$				& ${\rm AU}$		& Earth orbit semimajor axis \\
$e_E$		& $0.016751040$				& ---				& Earth orbit eccentricity \\
\hline
$\mu_M$		& $4.28280\times10^{4}$		& ${\rm km^3/s^2}$	& Gravitational parameter of Mars \\
$a_E$		& $1.523688399$				& ${\rm AU}$		& Mars orbit semimajor axis \\
$e_E$		& $0.093418671$				& ---				& Mars orbit eccentricity \\
\hline
\end{tabular}
\end{table}
\begin{table}[h!]
\caption{Bitangential transfers and Hohmann transfer.}
\label{tab:Hcases} 
\centering
\begin{tabular}{ccccccc}
\hline
Case	& Earth	& Mars 	& $\Delta V_1$ (km/s)	& $\Delta V_{2,\infty}$ (km/s)	& $\Delta V$ (km/s) & $\Delta t$ (days) \\
\hline
H1		& @P	& @P	& 2.179					& 3.388					& 5.568 			& 234 \\
H2		& @P	& @A 	& 3.398					& 2.090					& 5.488				& 278 \\
H3		& @A	& @P	& 2.414					& 3.163					& 5.577				& 239 \\
H4		& @A	& @A	& 3.629					& 1.881					& 5.510				& 283 \\
\hline
\end{tabular}
\end{table}

\end{document}